\documentclass[twocolumn,pra,aps,reprint,superscriptaddress,nofootinbib]{revtex4}
\usepackage{graphicx}
\usepackage{dcolumn}
\usepackage{bm}
\usepackage{bbm}
\usepackage[utf8]{inputenc}
\usepackage{amssymb}
\usepackage{amsmath}
\usepackage{bbold}
\usepackage{pstricks}
\usepackage{xcolor}
\usepackage{slashed}
\usepackage{braket}
\usepackage{hyperref}
\usepackage{natbib}
\usepackage{float}
\usepackage{todonotes}
\usepackage{mathrsfs}

\usepackage{siunitx}

\renewcommand{\Re}{\operatorname{Re}}
\renewcommand{\Im}{\operatorname{Im}}
\newcommand{\Tr}{\operatorname{Tr}}

\graphicspath{ {./figures/}   }
\newcommand{\fh}[1]{{\color{red}#1}}
\renewcommand{\fh}[1]{#1}

\begin{document}

\title{Implementing quantum electrodynamics with ultracold atomic systems}

\author{V. Kasper}
\email[]{kasper@physics.harvard.edu}
\affiliation{Institut f\"{u}r Theoretische Physik, Universit\"{a}t
Heidelberg, Philosophenweg 16,
69120 Heidelberg, Germany}
\affiliation{ Department of Physics, Harvard University, Cambridge, Massachusetts 02138, United States of America}

\author{F. Hebenstreit}
\affiliation{Albert Einstein Center, Institut f\"{u}r Theoretische Physik,
Universit\"{a}t Bern,
Sidlerstrasse 5, 3012 Bern, Switzerland}

\author{F. Jendrzejewski}
\author{M. K. Oberthaler}
\affiliation{Kirchhoff Institut für Physik, Universit\"{a}t Heidelberg, Im
Neuenheimer Feld 227,
69120 Heidelberg, Germany}

\author{J. Berges}
\affiliation{Institut f\"{u}r Theoretische Physik, Universit\"{a}t
Heidelberg, Philosophenweg 16,
69120 Heidelberg, Germany}


\begin{abstract}
We discuss the experimental engineering of model systems for the description
of QED in one spatial dimension via a mixture of bosonic $^{23}$Na and
fermionic $^6$Li atoms. The local gauge symmetry is realized in an optical
superlattice, using heteronuclear boson-fermion spin-changing interactions
which preserve the total spin in every local collision. We consider a large
number of bosons residing in the coherent state of a Bose-Einstein
condensate on each link between the fermion lattice sites, such that the
behavior of lattice QED in the continuum limit can be recovered. The
discussion about the range of possible experimental parameters builds, in
particular, upon experiences with related setups of fermions interacting
with coherent samples of bosonic atoms. We determine the atomic system's
parameters required for the description of fundamental QED processes, such
as Schwinger pair production and string breaking. This is achieved by benchmark calculations of the atomic system and of QED itself using functional integral techniques. Our results demonstrate that the dynamics of one-dimensional QED may be realized with ultracold atoms using state-of-the-art experimental
resources. The experimental setup proposed may provide a unique access to
longstanding open questions for which classical computational methods are no
longer applicable.
\end{abstract}


\maketitle

\section{Introduction}

The experimental engineering of atomic model systems for the description of dynamical gauge fields represents a major challenge with most important applications.
Fundamental gauge fields mediate the strong and electroweak forces between matter in the standard model of particle physics, where the photon of quantum electrodynamics (QED) is a most prominent representative~\cite{Weinberg:1995mt}.
Gauge fields can also arise as emerging degrees of freedom in strongly correlated condensed matter systems such as related to the quantum Hall effect~\cite{2008RvMP...80.1083N} or effective theories of spin liquids~\cite{2004PhRvB..70u4437H}. 
Most pressing questions concern the real-time evolution of strongly interacting gauge fields coupled to fermionic matter (QCD), such as realized during the early stages of our universe and explored in collisions of ultrarelativistic nuclei at giant laboratory facilities~\cite{Berges:2012ks}.

The complex many-body dynamics of gauge fields is often very difficult to
study in the original systems both experimentally and theoretically. For
instance, in a heavy-ion collision most experimental observables give
information only about the integrated space-time evolution of the system.
Its theoretical description is complicated by the fact that ab initio
computer simulations of the real-time dynamics can only be achieved in
certain limiting cases because of the so-called
sign-problem~\cite{2014PhRvD..89g4011B}. Here, the experimental engineering
of atomic quantum simulators appears as an attractive alternative, which may
provide a unique access to longstanding open questions~\cite{2013AnP...525..777W,0034-4885-79-1-014401,Tagliacozzo-2013}.

Compact table-top experiments with ultracold atoms are rather easily accessible and provide a very flexible testbed, with tunable interactions or reduced dimensionality by shaping the confining optical potential~\cite{Bloch:2008zzb,UBHD-67314892}. Since setups employing ultracold quantum gases can be largely isolated from the environment, they offer the possibility to study fundamental aspects such as the unitary real-time evolution of systems with engineered Hamiltonians to high accuracy~\cite{Schweigler-2015}.

The realization of external static gauge fields has been achieved in many impressive experiments with ultracold atoms, such as in
Refs.~\cite{Lin-2009,Struck-2013}. In contrast, no experimental implementation of gauge fields as dynamical degrees of freedom, as in QED or
QCD, in an atomic setup has been achieved yet, although it has been theoretically discussed in \cite{2013AnP...525..777W,0034-4885-79-1-014401,Tagliacozzo-2013}. 
The presence of dynamical gauge fields in the description of a physical system reflects an underlying local symmetry, whose space-time dependent transformation properties significantly constrains the quantum dynamics allowed. 
Implementing such a gauge symmetry for a system of bosonic and fermionic atoms can lead to involved constructions, which often rely on higher-order processes that are challenging to control experimentally.

Further experimental progress can be facilitated with the identification and implementation of simple gauge field theory examples. 
Abelian gauge symmetries, such as the local $U(1)$ symmetry of QED, clearly stand out in this respect, in particular, if they are implemented in one spatial dimension. 
Abelian gauge theories \fh{in the continuum} are simpler than non-Abelian theories such as QCD because of the absence of self-interactions of the gauge bosons. 
In QED the photon interacts only via processes involving electrons and positrons, while the gluons in QCD directly interact with each other.
\fh{In a lattice-discretized theory in more than one spatial dimension even QED magnetic fields appear as ring exchange interactions which makes a possible experimental implementation via effective interactions \cite{PhysRevLett.107.275301,2013PhRvL.110l5303B}, higher-order perturbative processes \cite{2013PhRvA..88b3617Z} or ancillary degrees of freedom \cite{zohar2016digital1,zohar2016digital2} 
involved.
In one spatial dimension, however, no QED magnetic field exists.}
This dramatically simplifies possible descriptions of the interaction of the remaining electric field with the fermions. 
In this case, the interaction terms respecting the local gauge symmetry can be realized using heteronuclear boson-fermion spin-changing interactions which preserve the total spin in every local collision~\cite{2013PhRvA..88b3617Z,2015arXiv150601238K}.
\fh{While a Jordan-Wigner transformation can be used to express the fermions as quantum spins, our construction does not rely on this mapping but directly simulates the fermionic degrees of freedom.
This is particularly important in view of experimental implementations of gauge theories in dimensions larger than one, for which the Jordan-Wigner mapping is less useful.} 
Despite the reduced complexity, the Abelian gauge theory setup in (1 + 1)
space-time dimensions still offers a rich phenomenology, including important
dynamical strong-field phenomena such
as Schwinger pair production~\cite{Szpak-2012,2015arXiv150601238K} and
string breaking~\cite{2012PhRvL.109q5302B}, which are highly relevant for
many systems also in more than one spatial dimension.

Very interesting and detailed suggestions have been made to realize gauge
field dynamics in atomic systems, where many proposals concentrate on
quantum link models~\cite{2013AnP...525..777W}. In these models the
gauge fields are regularized using quantum link variables which have a
finite-dimensional link Hilbert space, and whose dimension is determined by
the number of bosonic atoms residing on a given link between fermionic atoms
in an optical lattice. Since the Hilbert space of a quantum link model is
finite, the mapping to atomic systems is in general facilitated.
Many ground-breaking investigations have been performed using a small number
of bosons per
link~\cite{2012PhRvL.109q5302B,2012PhRvL.109l5302Z,2013PhRvL.110l5303B,
Tagliacozzo-2013,2013PhRvA..88b3617Z}. A low-dimensional Hilbert space
also allows one to achieve theoretical estimates based on diagonalization or
tensor network techniques~\cite{Kuehn-2014,Pichler-2015}. Since the Hilbert
space of QED or QCD itself is infinite-dimensional, the Hamiltonian
formulation of the original gauge field theory on a spatial
lattice~\cite{1975PhRvD..11..395K} can be recovered for a sufficiently large
number of bosons\footnote{The mapping becomes more involved if only a small
number of bosons per link is employed. In this case, the quantum fields of
the original gauge theory can arise as low-energy effective degrees of
freedom of the theory of quantum links after dimensional reduction. More
precisely, the quantum fields of a gauge theory in D dimensions are obtained
from a (D + 1)-dimensional theory of quantum links. To recover
one-dimensional QED or QCD would, therefore, require a two-dimensional
quantum link setup where the extra dimension could also be implemented with
the help of internal degrees of freedom~\cite{2013AnP...525..777W}.}.

In this work we follow Ref.~\cite{2015arXiv150601238K} and consider
a mixture of bosonic $^{23}$Na and fermionic $^6$Li atoms in a
one-dimensional optical superlattice. We concentrate on the regime with a
large number of bosons residing on each link, such that the results of the original lattice gauge
theory in the continuum limit are recovered. Our discussion about the range
of possible atomic system's parameters builds, in particular, upon
experiences with related experimental setups of fermions interacting with
coherent samples of bosonic
atoms~\cite{Strobel-2014,Gross-2010,Hume-2013,Schuster-2012,Scelle2013a,Gross-2011}. The other important ingredient of our investigation concerns
benchmark calculations of the atomic system and of the original gauge
theory. Since exact diagonalization techniques are no longer applicable in
this case, we employ powerful functional integral (FI)
techniques~\cite{2013PhRvD..87j5006H,2013PhRvL.111t1601H,2014PhRvD..90b5016K,2015arXiv150601238K,Gelfand:2016prm}. They allow us to do ab initio
calculations in an important range of strong-field phenomena. Reproducing
these benchmark results with future experimental realizations will be a
crucial milestone, before new regimes can be explored that are no longer
accessible with classical computational methods.

This publication is organized as follows.
In section~\ref{sec:II} we briefly review QED in 1+1 space-time dimensions,
i.e.~the massive Schwinger model. We employ a lattice discretization with
staggered fermions to connect the gauge theory to an atomic model in an
optical superlattice with angular momentum conserving scattering processes.
In section~\ref{sec:III} we discuss in detail a possible experimental
implementation of the Schwinger model in a mixture of bosonic and fermionic
atoms, where gauge invariance requires a correlated hopping of the staggered
fermions with the Schwinger bosons residing on the links. While the basic
discussion follows to a large extent Ref.~\cite{2013PhRvA..88b3617Z}, we
concentrate on the implementation in specific systems with given
experimental parameters to ensure the relevant separation of scales that is
required to suppress contributions from gauge-symmetry violating states.
Moreover, we employ species-dependent lattices to separate the bosonic and
fermionic degrees of freedom in order to simplify the experimental
realization. In that section we also translate the basic quantities of the
cold atom system to the fundamental parameters of the corresponding lattice
gauge theory. A set of viable parameters to be employed in an upcoming
experiment is presented in section~\ref{sec:IV}.
The later sections are devoted to benchmark calculations in order to
demonstrate that relevant QED processes can indeed be described using the
available experimental parameters of the atomic setup.
In section~\ref{sec:V} we review the functional integral approach and derive
equations of motion for the cold atom system.
This method allows us to accurately describe the nonequilibrium dynamics of
coherent bosonic fields coupled to fermions from first principles.
In section \ref{sec:VI} we study the real-time dynamics of Schwinger pair
production and string breaking in the cold atom system. This section is an
extension of the pair-production results of Ref.~\cite{2015arXiv150601238K}
to the new parameter sets established in this work, and to the phenomenon of
string breaking that has not been considered in the large boson number
regime of the atomic setup before.
We present the dynamics of various experimentally accessible observables and
discuss the accuracy with which QED can be represented in practice
by a finite atomic system. We conclude and give an outlook in section \ref{sec:VII}.

\section{The Schwinger model revisited \label{sec:II}}

Quantum electrodynamics for massless fermions in one spatial dimension (Schwinger model) is an exactly solvable field theory \cite{1962PhRv..128.2425S}.
On the other hand, no analytic solution is known for massive fermions (massive Schwinger model) \cite{1976AnPhy.101..239C}.
From a phenomenological point of view, a particular interest in this model stems from the fact that it shares several characteristic aspects with the theory of strong interactions (QCD) such as spontaneous chiral symmetry breaking or dynamical string breaking~(see e.g.~Refs.~\cite{PhysRevD.44.257,2000PhRvD..62i4504M,PhysRevD.87.077501,PhysRevD.89.074053}).

Nonperturbative studies of the massive Schwinger model are typically based on a lattice discretization of the continuum theory.
\fh{The construction of a hermitean, local and translation-invariant lattice theory of fermions necessarily entails the appearence of unphysical degrees of freedom, the so-called fermion doublers \cite{Nielsen:1980rz}.
One possibility to resolve this problem is by making the spurious doubler modes heavy via a Wilson term \cite{PhysRevD.10.2445} and we refer to Refs.~\cite{2013PhRvD..87j5006H,2013PhRvL.111t1601H,2014PhRvD..90d5034H} for numerical studies using this approach.
In this work, we employ the alternative staggered fermion discretization where the Dirac spinor is decomposed in such a way that the doubler modes can be disregarded as they decouple.
As a consequence, the} particle and antiparticle components reside on neighbouring lattice sites \cite{1975PhRvD..11..395K}.
The Hamiltonian of the theory is given by 
\begin{align}
\label{eq:KS_Hamiltonian}
 H_{\text{KS}}&=\frac{a_S}{2}\sum_n E^2_{n} + M \sum_n (-1)^n \psi^{\dagger}_n \psi_n \notag \\
 &-\frac{i}{2a_S}\sum_n \left( \psi^{\dagger}_{n} U_{n} \psi_{n+1} - h.c.\right) \ ,
\end{align}
where $a_S$ denotes the lattice spacing and $M$ the fermion mass \cite{1962PhRv..128.2425S,1975PhRvD..11..395K} and we choose to work in natural units ($\hbar=c=1$).
Here, the staggered fermion field operator $\psi_n$, which resides on lattice sites $n$, fulfills the canonical anti-commutation relation $\{\psi_n,\psi^\dagger_m\}=\delta_{nm}$.
\fh{The fermionic charge operator is defined as
\begin{align}
 \mathcal{Q}_n=\psi^\dagger_n\psi_n + \frac{(-1)^n - 1}{2} \ .
\end{align}
Accordingly, the presence of a fermion at an even site is interpreted as particle ($\mathcal{Q}_n=+1$) whereas the absence of a fermion at an odd site is interpreted as antiparticle ($\mathcal{Q}_n=-1$).}
The \fh{unitary} link operator $U_n$ and \fh{the} electric field operator $E_n$ act between neighboring lattice sites $n$ and $n+1$ and obey the commutation relations
\begin{subequations}
\begin{align}
 \label{eq:comm1}
 [U_n,U^\dagger_m]&=0 \ , \\
 [E_n,U_m]&=g\delta_{nm}U_m \ ,
 \label{eq:comm2}
\end{align}
\end{subequations}
where $g$ denotes the gauge coupling.
This algebra entails an infinite-dimensional local Hilbert space.
For the $U(1)$ gauge theory, the Gauss's law operator
\begin{equation}
 G_n=E_n-E_{n-1}-g\fh{\mathcal{Q}_n} 
 \label{eq:Gn}
\end{equation} 
generates local gauge transformations and commutes with the Hamiltonian $[H_{\text{KS}},G_n]=0$.

\begin{figure}[t]
 \includegraphics[width=\columnwidth]{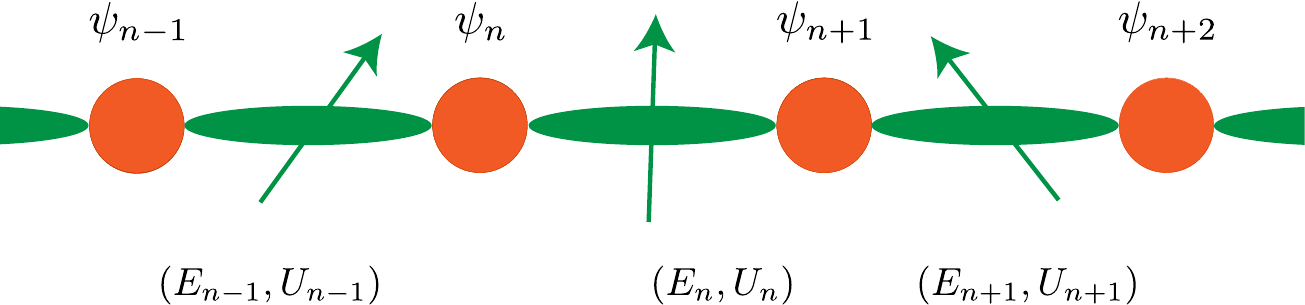}
 \caption{\fh{Schematic representation of the Hamiltonian \eqref{eq:KS_Hamiltonian}: The fermionic modes $\psi_n$ on each lattice site (orange circles) interact with $U(1)$ variables at each link $(E_n,U_n)$, that connect neighbouring lattice sites (green clouds).}}
 \label{fig:0}
\end{figure}

Quantum link models have been proposed as an alternative formulation of gauge theories for which finite-dimensional representations of the link algebra exist \cite{HornQuantumLink, 1990NuPhB.338..647O, 1997NuPhB.492..455C}.
Recently, the prospect of constructing quantum simulators for gauge theories has boosted the interest in these models as their implementation in atomic systems could be greatly facilitated.
In this approach, the electric field operator is identified with the $z$-component of the quantum spin operator $L_{z,n}$,
\begin{equation}
E_n \to gL_{z,n} \,,
\end{equation}
whereas the link operators are regarded as raising and lowering operators
\begin{subequations}
\begin{align}
 U_n&\to[\ell(\ell+1)]^{-1/2}L_{+,n} \, , \\
 U^\dagger_n&\to [\ell(\ell+1)]^{-1/2}L_{-,n}\, , 
\end{align} 
\end{subequations}
with $L_{\pm,n}=L_{x,n}\pm i L_{y,n}$.
The quantum spin operators fulfill the angular momentum algebra $[L_{i,n},L_{j,m}]=i\delta_{nm}\epsilon_{ijk}L_{k,n}$ and $\ell$ denotes the spin magnitude.
Consequently, the commutation relation \eqref{eq:comm2} is identically fulfilled, whereas the commutation relation \eqref{eq:comm1} is no longer valid if $\ell$ is kept finite, but replaced by $[U_n,U_m^\dagger]=2 \delta_{nm}E_m/[g \ell(\ell+1)]$ which goes to zero only as $\ell \rightarrow \infty$.
\fh{Only in this limit the unitarity of the link operator $U_n$ is restored again.}
However, it is central for the whole construction that a finite $\ell$ does not affect gauge invariance as generated by the Gauss's law operator 
\begin{equation}
G_n \to L_{z,n} - L_{z,n-1} - \fh{\mathcal{Q}_n}
\end{equation}
with the Hamiltonian of the quantum link Schwinger model
\begin{align}
\label{eq:QL_Hamiltonian}
 H_{\text{QL}}&=\frac{g^2 a_S}{2}\sum_n L^2_{z,n} + M \sum_n (-1)^n \psi^{\dagger}_n \psi_n \notag \\
 &-\frac{i}{2a_S\sqrt{\ell(\ell+1)}}\sum_n \left( \psi^{\dagger}_{n} L_{n,+} \psi_{n+1} - h.c.\right) \ .
\end{align}
The finite-dimensional representation of the angular momentum algebra makes its implementation in systems of ultracold atoms feasible. For representations with small $\ell$, numerical methods based on diagonalization or tensor network states provide valuable information about static and dynamic properties~\cite{2014PhRvL.112t1601R,2013JHEP...11..158B,Kuehn-2014,Buyens-2014,Pichler-2015}. Of course, it is an important question to understand the connection between the finite-dimensional representation of cold atom 
gauge theories and the infinite-dimensional representation corresponding to QED. In Ref.~\cite{2015arXiv150601238K} the large-$\ell$ regime and the convergence to $\ell \to \infty$ results, i.e.~QED itself, has been established using powerful functional integral (FI) techniques for strong-field phenomena. In section~\ref{sec:V} we describe the FI approach and apply it to obtain benchmark results for the Hamiltonian (\ref{eq:QL_Hamiltonian}) using parameter sets motivated by possible experimental realizations.

\section{Experimental realization \label{sec:III}}

Our starting point for the realization of the $U(1)$ 
gauge theory coupled to fermionic matter in an 
ultracold atom experiment is a genuine interacting 
gas of fermionic and bosonic atoms~\cite{0034-4885-79-1-014401}. 
To facilitate the connection with the experiments we reintroduce $\hbar$ where appropriate.
The bosons $\phi_\alpha(\mathbf{x})$ and fermions $\psi_\alpha(\mathbf{x})$ fulfill the canonical commutator and anti-commutator relations, respectively,
\begin{equation}
 [\phi_{\alpha}(\mathbf{x}_1),\phi^\dagger_{\beta}(\mathbf{x}_2)]= \{\psi_{\alpha}(\mathbf{x}_1),\psi^\dagger_{\beta}(\mathbf{x}_2)\} = \delta_{\alpha\beta}\delta(\mathbf{x}_1-\mathbf{x}_2) \ . 
\end{equation}
Here, the greek labels $\alpha, \beta$ denote magnetic hyperfine states of the atoms.
The particles are confined by external potentials and interact via inter- and intra-species scattering processes.  
The corresponding Hamiltonian consists of three parts, $H = H_T + H_V + H_I$. The kinetic part, $H_T$, describes the movement of the atoms,
\begin{align}
H_T & \ =\ \frac{\hbar^2 }{2M_b}  \int d^3 \mathbf{x}  \sum_{\alpha} |\nabla \phi_{\alpha}(\mathbf{x})|^2  \notag \\
     &\ +\ \frac{\hbar^2}{2M_f}  \int d^3 \mathbf{x}  \sum_{\alpha} |\nabla \psi_{\alpha}(\mathbf{x})|^2 \ 
\label{eq:H_kin}
\end{align} 
with masses $M_b$ and $M_f$. The potential energy contribution, $H_V$, is determined by the external potentials
according to
\begin{align}
H_V & \ =\ \int d^3 \mathbf{x}  \sum_{\alpha} V^b_{\alpha}(\mathbf{x}) \phi^{\dagger}_{\alpha}(\mathbf{x}) \phi_{\alpha}(\mathbf{x})  \notag \\
    & \ +\ \int d^3 \mathbf{x}  \sum_{\alpha} V^f_{\alpha}(\mathbf{x}) \psi^{\dagger}_{\alpha}(\mathbf{x}) \psi_{\alpha}(\mathbf{x}) \ ,
\label{eq:H_Pot}
\end{align}
whereas the atomic scattering processes are described by
\begin{align}
H_I & \ =\  \frac{1}{2}\int d^3 \mathbf{x} \sum_{\alpha\beta\gamma\delta}g^b_{\alpha\beta\gamma\delta}  \,
 \phi^{\dagger}_{\alpha}(\mathbf{x}) \phi^{\dagger}_{\beta}(\mathbf{x}) \phi_{\delta}(\mathbf{x}) \phi_{\gamma}(\mathbf{x}) \notag \\
    & \ +\ \frac{1}{2}\int d^3 \mathbf{x} \sum_{\alpha\beta\gamma\delta} g^f_{\alpha\beta\gamma\delta}  \,  
 \psi^{\dagger}_{\alpha}(\mathbf{x}) \psi^{\dagger}_{\beta}(\mathbf{x}) \psi_{\delta}(\mathbf{x}) \psi_{\gamma}(\mathbf{x}) \notag \\
    & \ +\ \frac{1}{2}\int d^3 \mathbf{x} \sum_{\alpha\beta\gamma\delta} g^{bf}_{\alpha\beta\gamma\delta}
 \psi^{\dagger}_{\alpha}(\mathbf{x}) \phi^{\dagger}_{\beta}(\mathbf{x}) \phi_{\delta}(\mathbf{x}) \psi_{\gamma}(\mathbf{x})
  \, .
\, \label{eq:FullBFInteraction}
\end{align}
The coupling constants are determined by the scattering lengths of the inter- and intra-species scattering processes.
Throughout this work, we indicate purely bosonic terms by 
the superscript $b$, fermionic terms by the superscript $f$ and boson-fermion interactions by the superscript $bf$.

\fh{In the following, we describe in detail all the steps that are required so that the gas of fermionic and bosonic atoms in three spatial dimensions \eqref{eq:H_kin} -- \eqref{eq:FullBFInteraction} behaves according to the Hamiltonian \eqref{eq:QL_Hamiltonian} in one spatial dimension.
To this end, we show how to reduce the dimensionality from three to one dimensions and how to construct the staggered lattice for fermions.
Afterwards, we describe how to select only those interactions from \eqref{eq:FullBFInteraction} that correspond to the gauge-invariant interactions in \eqref{eq:QL_Hamiltonian}.}

\subsection{One-dimensional staggered geometry}

\begin{figure}[t]
 \includegraphics[width=\columnwidth]{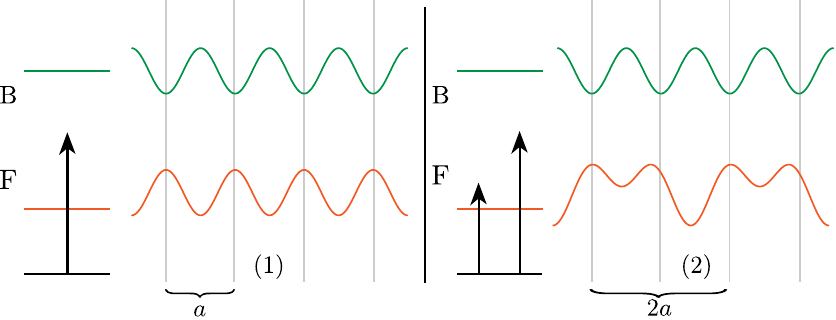}
 \caption{
 Realization of the staggered lattice structure. The black arrows indicate the dipole lasers which are used in the experimental setup.
 (1) Blue/red detuning for fermions/bosons generates phase-shifted optical potentials for bosons and fermions with a lattice period $a$.
 (2) The superposition of the lattice with period $a$ by a superlattice of period $2a$ generates the staggered structure for the fermions.}
 \label{fig:1}
\end{figure}

The basic ingredient for realizing a one-dimensional lattice structure with lattice constant $a$ is an optical lattice with tight radial confinement.
Employing a laser frequency which is blue detuned for fermions and red detuned for bosons allows us to place a mesoscopic bosonic gas on the links between the fermionic lattice sites, as indicated in the left graph of Fig.~\ref{fig:1}.
In fact, the potential energy contributions \eqref{eq:H_Pot} can be split into an axial and radial part,
\begin{align}
 V^{s}_{\alpha}(\mathbf{x}) = V_{\parallel,\alpha}^{s}(x) + V^{s}_{\perp,\alpha}e^{-(y^2+z^2)/l^2_{s,\alpha}} \, .
\end{align}
Here, $l_{s,\alpha}$ denotes the radial confinement length scale, where
we distinguish bosons and fermions by the superscript $s\in\{b,f\}$.
Owing to tight radial confinement, the three-dimensional system is effectively rendered one-dimensional and we employ the product
form
\begin{subequations}
\label{eq:Reduction}
\begin{align}
\phi_{\alpha}(\mathbf{x}) &= \varphi_{b}(y)\varphi_{b}(z) \phi_{\alpha}(x) \, ,\\
\psi_{\alpha}(\mathbf{x}) &= \varphi_{f}(y)\varphi_{f}(z) \psi_{\alpha}(x) \, , 
\end{align} 
\end{subequations}
where $\varphi_{s}(y)$ and $\varphi_{s}(z)$ are the ground state wave functions in the $y$ and $z$ directions, respectively.
We assume that these states are independent of the magnetic quantum number.

To generate a staggered structure for the fermions, the original optical lattice with period $a$ is superimposed by an optical superlattice with period $2a$, as indicated in the right graph of Fig.~\ref{fig:1}.
Owing to the fact that the frequency of the second laser is tuned closer to resonance with respect to the fermions than to the bosons, the second lattice does practically not affect the bosonic degrees of freedom.
Disregarding the effect of overall confinement in the axial direction, the axial part of the potential is then given by
\begin{subequations}
\label{eq:LatticePotential}
\begin{align}
 V^{b}_{\parallel, \alpha}(x) &=  V^{b}_{1,\alpha} \cos^2\Big(\frac{2\pi x}{a}\Big) \ , \\
 V^{f}_{\parallel, \alpha}(x) &=  V^{f}_{1,\alpha} \sin^2\Big(\frac{2\pi x}{a}\Big) + V^{f}_{2,\alpha} \cos^2 \left(\frac{\pi x}{a}\right) \ . 
\end{align}
\end{subequations}
The amplitudes $V^{s}_{i,\alpha}$, $i=1,2$ are determined by the AC Stark shift of the corresponding magnetic substates $\alpha$.
The tunneling of bosons between adjacent sites is suppressed by tuning of the laser amplitude.
As already noted in Ref.\mbox{$\, $}\cite{2013PhRvA..88b3617Z}, such a construction is spin-independent and, hence, distinct from e.g.~Ref.\mbox{$\, $}\cite{2012PhRvL.109q5302B}.

In the following, it is useful to switch to a representation in terms of localized Wannier functions.
To this end, we first consider the bosonic degrees of freedom and focus on the two lowest energy bands.
By tuning of the laser amplitude, we may choose $V^f_{i,\alpha}$ such that two Wannier functions $w^{f}_{\alpha,n'p}(x)$ are obtained which are sufficiently localized in the left ($p=L$) and right ($p=R$) minimum of the elementary cell $n'\in\{0,\ldots,N-1\}$ with positive integer $N$ of the optical lattice, respectively.
The corresponding expansion of the fermionic field operator reads
\begin{align}
 \psi_{\alpha}(x) = \sum_{n',p} w^f_{\alpha, n' p}(x)\, \psi_{\alpha,n'p} \ .
\end{align}
We note that the total number of bosonic/fermionic lattice sites is $2N$ and we will label them by $n\in\{0,\ldots,2N-1\}$.
Similarly, we may expand the bosonic field operators, $\phi_{\alpha}(x) = \sum_{n',p} w^b_{\alpha, n' p}(x) \phi_{\alpha,n'p}$, where the Wannier functions $w^{b}_{\alpha,n'p}(x)$ are again localized in the two minima of the elementary cell. 
In fact, the structure of the superlattice suggests the definition 
\begin{subequations}
\begin{alignat}{4}
&\psi_{2n',\alpha} \equiv \psi_{\alpha,n'L}&\quad , \quad &\psi_{2n'+1, \alpha} \equiv \psi_{\alpha,n'R} \, , \\
&\phi_{2n',\alpha} \equiv \phi_{\alpha,n'L}&\quad , \quad &\phi_{2n'+1, \alpha} \equiv \phi_{\alpha,n'R}    \, . 
\end{alignat}
\end{subequations}
We note that the kinetic energy contributions \eqref{eq:H_kin} are suppressed since the Wannier functions that correspond to the different minima in the optical lattice do not have a sizable overlap (cf.~also section~\ref{sec:IV}).

\subsection{Angular momentum conservation}

In the previous section, we reviewed how the potential energy \eqref{eq:H_Pot} can be used to generate the staggered lattice structure.
Moreover, is was noted that the kinetic energy \eqref{eq:H_kin} is suppressed due to the localization of the Wannier functions at the potential minima.
To ensure local gauge invariance and create dynamics in the cold atom system, we have to tune the interaction Hamiltonian \eqref{eq:FullBFInteraction} such that only a selection of terms contributes. In the following,
we discuss all interaction terms in more detail and explain the connection between the various scattering lengths and coupling constants $g^{s}_{\alpha \beta \gamma \delta}$ for $s\in\{b,f,bf\}$.
For pedagogical reasons, we discuss the construction in free space first and take into account the lattice later.

We suppose that the inter- and intra-species interactions of bosons and fermions are local and conserve angular momentum.
Specifically, we consider bosonic degrees of freedom with spin $f_b=1$ and fermionic degrees of freedom with spin $f_f=1/2$.
Therefore, the two-particle potentials are given by
\begin{align}
 V^s(\mathbf{x}_1,\mathbf{x}_2)=\delta(\mathbf{x}_1-\mathbf{x}_2)\sum_{\mathcal{F}_s}g_{s,\mathcal{F}_s}\mathbf{P}_{\mathcal{F}_s} \ ,
\end{align}
where the total spin can take the values $\mathcal{F}_b\in\{0,2\}$, $\mathcal{F}_f\in\{0,1\}$ and $\mathcal{F}_{bf}\in\{1/2,3/2\}$ \cite{Kawaguchi2012253}.
The interaction strengths $g_{s,\mathcal{F}_s}$ are related to the s-wave scattering lengths $a_{s,\mathcal{F}_s}$ via
\begin{align}
 g_{s,\mathcal{F}_s}=\frac{2\pi \hbar^2 a_{s,\mathcal{F}_s}}{M_{r,s}} \ .
\end{align}
Here $M_{r,s}$ denotes the reduced mass of the two scattering partners.
In general, the projector $\mathbf{P}_{\mathcal{F}}$ for two particles with individual spins $f_1$ and $f_2$ on the subspace with total spin $\mathcal{F}$ can be written as
\begin{equation}
 \mathbf{P}_{\mathcal{F}}=\sum_{M}\ket{f_1,f_2;\mathcal{F},M}\bra{f_1,f_2;\mathcal{F},M} \ , 
\end{equation}
where $M\in\{-\mathcal{F},-\mathcal{F}+1,\ldots,\mathcal{F}-1,\mathcal{F}\}$ are the possible magnetic quantum numbers. We may relate the interaction strengths $g_{s,\mathcal{F}_s}$ to the constants appearing in the interaction part of the Hamiltonian \eqref{eq:FullBFInteraction} according to
\begin{align}
 \label{eq:coupling_CG}
 g^{s}_{\alpha\beta\gamma\delta} &=  \sum_{\mathcal{F}_s} \sum_{M}g_{s,\mathcal{F}_s}\braket{f_1;\alpha;f_2;\beta|f_1,f_2;\mathcal{F}_s,M} \notag \\ 
 &\times  \braket{f_1,f_2;\mathcal{F}_s,M|f_1;\gamma ;f_2;\delta} \, .
\end{align}
Here, $\braket{f_1;\alpha;f_2;\beta|f_1,f_2;\mathcal{F}_s,M}$ are the Clebsch-Gordan coefficients for coupling the individual spins $f_1$ and $f_2$ to the total spin $\mathcal{F}_s$.
Specifically, we have $f_1=f_2=1$ for boson-boson interactions ($s=b$), $f_1=f_2=1/2$ for fermion-fermion interaction ($s=f$) and $f_1=1/2$, $f_2=1$ for boson-fermion interaction ($s=bf$).

Following along the lines of the previous section, we reduce the three-dimensional system to a setup with effectively one spatial dimension and expand the field operators in terms of Wannier functions.
Using a compact notation, where $\mathbf{n}=(n_1,n_2,n_3,n_4)$ denotes the site indices and $\boldsymbol{\mu}=(\alpha,\beta,\gamma,\delta)$ the magnetic quantum numbers, we can write for the interaction part of the Hamiltonian:
\begin{align}
H_I &= \frac{1}{2}\sum_{\mathbf{n},\boldsymbol{\mu}}  U^b_{\mathbf{n}} {g}^b_{\boldsymbol{\mu}} 
\phi^{\dagger}_{n_1,\alpha} \phi^{\dagger}_{n_2,\beta} \phi_{n_4,\delta} \phi_{n_3,\gamma}
\notag \\
&+\frac{1}{2} \sum_{\mathbf{n},\boldsymbol{\mu}} U^{f}_{\mathbf{n}}{g}^{f}_{\boldsymbol{\mu}} 
\psi^{\dagger}_{n_1,\alpha} \psi^{\dagger}_{n_2,\beta} \psi_{n_4,\delta} \psi_{n_3,\gamma}
\notag \\
&+\frac{1}{2}\sum_{\mathbf{n},\boldsymbol{\mu}} U^{bf}_{\mathbf{n}}{g}^{bf}_{\boldsymbol{\mu}}
\psi^{\dagger}_{n_1,\alpha} \phi^{\dagger}_{n_2,\beta} \phi_{n_4,\delta} \psi_{n_3,\gamma}\, .
\label{eq:HI2}
\end{align}
The coupling constants $U^{s}_{\mathbf{n}}$ are determined by the dimensional reduction and the explicit form of the overlap integrals of the Wannier functions, as given in Appendix~\ref{sec:app_overlap}.

\begin{figure}[t]
 \includegraphics[width=0.9\columnwidth]{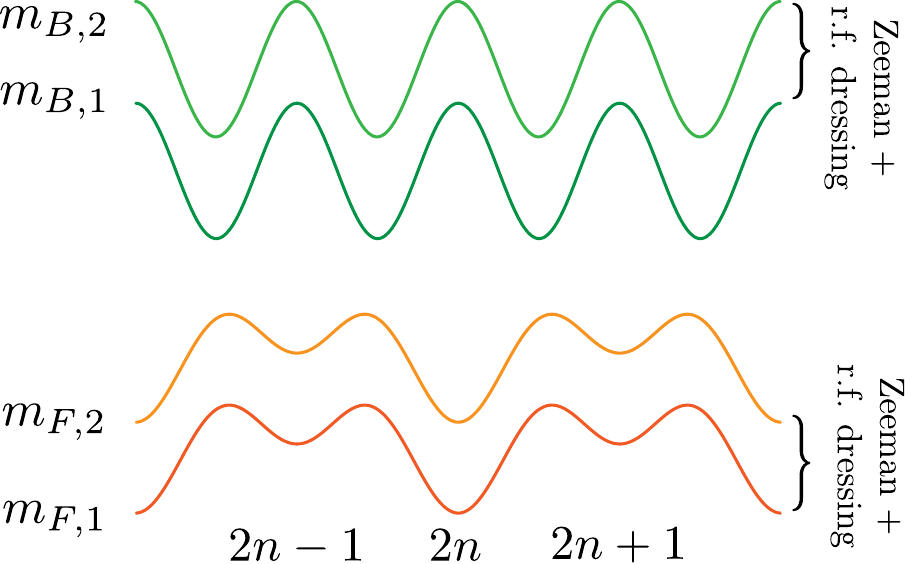}
 \caption{
 The application of a $B$-field and rf-dressing splits the superlattice for the individual magnetic substates of the bosons and fermions.}
 \label{fig:2}
\end{figure}

Based on this interaction Hamiltonian, we see that a plethora of possible interaction terms are generated in general.
In order to realize the Hamiltonian \eqref{eq:QL_Hamiltonian}, however, we have to guarantee that only specific terms contribute that respect the gauge symmetry.
To this end, we use the fact that the application of an appropriate magnetic field and radio-frequency (rf) dressing allows for a selection of a small number of relevant interaction terms, whereas all other contributions become suppressed.
We emphasize that this selection is achieved by the unequal shift of the bosonic and fermionic energy levels, as depicted in Fig.~\ref{fig:2}. 
Most notably, this procedure results in the bosonic spin exchange with the simultaneous fermion hopping that corresponds to the gauge-invariant interaction term in \eqref{eq:QL_Hamiltonian}.
We note that this selection process does not exclude elastic scattering terms, i.e.\ scattering processes without changing the individual spins of the atoms. 

\begin{figure}[t]
 \includegraphics[width=\columnwidth]{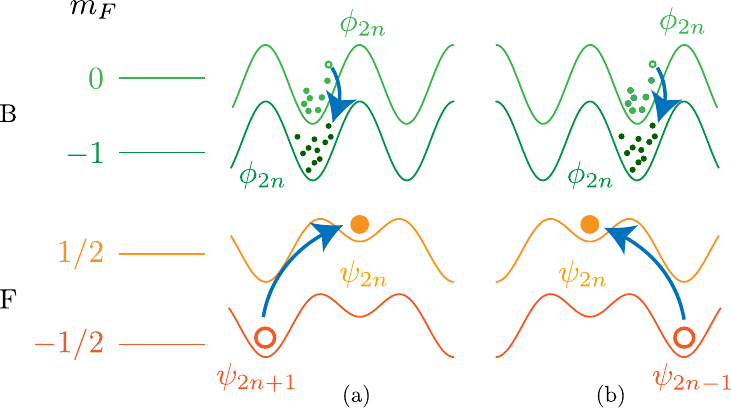}
 \caption{
 The selection procedure results in the correlated bosonic spin exchange with a fermionic hopping in the superlattice. 
 Note that the inverse process is allowed as well.}
 \label{fig:3}
\end{figure}

All bosonic states are prepared in $\alpha_b=\{-1,0\}$ 
states whereas the fermionic degrees of freedom are generated in the staggered configuration with $\alpha_{f}=1/2$ on even sites and $\alpha_f=-1/2$ on odd sites.
As a consequence, interactions including the $\alpha_b=1$ sector, which are allowed in principle, are suppressed at all times if initialized accordingly. We further elaborate on this issue in the following sections.

\subsection{Bosonic intra-species interactions}

In this section, we discuss the intra-species interaction terms of bosons in more detail.
Owing to localization in the optical lattice, only on-site interactions of bosons contribute, i.e.\ $U^{b}_{\mathbf{n}}\neq0$  for $\mathbf{n}=(n,n,n,n)$ and all others effectively vanish.
Accordingly, the relevant part of the purely bosonic term in the interaction Hamiltonian (\ref{eq:HI2}) is given by
\begin{align}
H^{b}_{I} &=  \frac{1}{2}\sum_{n,\boldsymbol{\mu}} U^{b}_{\mathbf{n}} g^{b}_{\boldsymbol{\mu}}\phi^{\dagger}_{n,\alpha} \phi^{\dagger}_{n,\beta} \phi_{n,\delta} \phi_{n,\gamma}\, , 
\label{eq:BosonInteraction}
\end{align}
where all entries of $\boldsymbol{\mu}=(\alpha,\beta,\gamma,\delta)$ may take values $\alpha_b\in\{-1,0\}$.
Again, we note that we disregard terms including the magnetic substates $\alpha_b=1$ which are excluded by the spin conservation if initialized accordingly. 
The interaction term \eqref{eq:BosonInteraction} reads
\begin{align}
 H^{b}_{I} = \frac{1}{2} \sum_{n} U^b_{\mathbf{n}}& \left(g_{b,2}\,\phi_{n,-1}^\dagger\phi_{n,-1}^\dagger\phi_{n,-1}\phi_{n,-1} \right.\notag \\
                                                  & \left.+ \frac{g_{b,0}+2g_{b,2}}{3} \phi_{n,0}^\dagger\phi_{n,0}^\dagger\phi_{n,0}\phi_{n,0}\right. \notag \\
                                                  & \left.+ 2g_{b,2}\,\phi_{n,0}^\dagger\phi_{n,-1}^\dagger\phi_{n,-1}\phi_{n,0}\right) \ ,
\end{align}
where the coupling constants result from \eqref{eq:coupling_CG} and we assumed that the overlap integrals $U^{b}_{\mathbf{n}}$ are the same for all terms.
For later convenience, we denote the bosonic degrees of freedom on even sites as $\phi_{2n,-1}\equiv d_{2n}$ and $\phi_{2n,0}\equiv b_{2n}$, whereas we interchange their role on odd sites such that $\phi_{2n+1,-1}\equiv b_{2n+1}$ and $\phi_{2n+1,0}\equiv d_{2n+1}$.
In fact, the bosons $b_n$ and $d_n$ can be understood as Schwinger bosons \cite{UBHD-9825719} with the identification
\begin{subequations}
\label{eq:Schwinger_bosons}
\begin{align}
 L_{+,n} &\ =\ b^{\dagger}_n d_n \ , \ L_{-,n} \ =\ d_n^{\dagger} b_n \, , \\
 L_{z,n} &\ =\ \frac{1}{2}(b^{\dagger}_n b_n - d^{\dagger}_n d_n) \, , 
\end{align}
\end{subequations}
which constitute a representation of the angular momentum algebra $[L_{i,n},L_{j,m}]=i\delta_{nm}\epsilon_{ijk}L_{k,n}$ with $L_{\pm,n}=L_{x,n}\pm iL_{y,n}$. 
The constraint $2\ell = b^{\dagger}_n b_n + d_n^{\dagger} d_n$ is fulfilled because the hopping of bosons between neighboring sites $n\to n\pm1$ is suppressed. 
The bosonic intra-species interaction Hamiltonian 
is then given by
\begin{align}
H^{b}_{I} &= \frac{g_{b,0}-g_{b,2}}{6}U_{\mathbf{n}}^b\sum_n L^2_{z,n}  + \Delta_{b,0} \sum_n (-1)^n L_{z,n}  \, ,
\end{align}
where we disregarded an irrelevant constant
and introduced the abbreviation $\Delta_{b,0} \equiv (2\ell-1)\frac{g_{b,0}-g_{b,2}}{6}U_{\mathbf{n}}^b$.

\subsection{Fermionic intra-species interaction term}\label{Sec:FermInteractions}

In this section, we discuss the intra-species interaction terms of fermions in more detail.
Again, only on-site interaction terms contribute owing to localization such that $U^{f}_{\mathbf{n}}\neq0$ only for $\mathbf{n}=(n,n,n,n)$.
Taking into account the Clebsch-Gordon coefficients, the 
purely fermionic term in the interaction Hamiltonian (\ref{eq:HI2})
can be reduced according to
\begin{align}
 \label{eq:fourfermi}
 H^{f}_{I} &=  \sum_{n} U^{f}_{\mathbf{n}} g_{f,0} \psi^{\dagger}_{n,1/2} \psi^{\dagger}_{n,-1/2} \psi_{n,-1/2} \psi_{n,1/2} \ .
\end{align}
In general, this four-fermion interaction term influences the dynamics.
However, the contribution can be written as a density-density interaction 
\begin{align}
 H^{f}_{I} &= \sum_{n}U^{f}_{\mathbf{n}} g_{f,0}\rho_{n,1/2} \rho_{n,-1/2} \, 
\label{eq:StaggeredInteraction}
\end{align}
between $\alpha_f=-1/2$ and $\alpha_f=1/2$ particles with density operators $\rho_{n,\pm1/2}=\psi^{\dagger}_{n,\pm1/2} \psi_{n,\pm1/2}$.
Restricting ourselves to an initial state $\ket{\Psi}$ with only $\alpha_{f}=1/2$ particles on even sites and only $\alpha_f=-1/2$ particles on odd sites, one immediately finds that $H^{f}_{I}\ket{\Psi}=0$. Consequently, this four-fermion interaction does not contribute to the time evolution due to an appropriate initial-state preparation.

\subsection{Inter-species interaction term}

Regarding the fermion-boson scattering contributions to the Hamiltonian (\ref{eq:HI2}), we have to consider both the spin exchange process as well as elastic scattering processes.
According to the interaction selection process described above, the spin exchange term that involves the correlated hopping of fermions and bosons is given by
\begin{align}
H^{bf}_{I_{se}} =\! \frac{1}{2}&\sum_{n'} U^{bf}_{\mathbf{n}} g^{bf}_{\boldsymbol{\mu}} \big.(\psi^{\dagger}_{2n,\alpha} \phi^{\dagger}_{2n,\beta} \phi_{2n,\delta} \psi_{2n+1,\gamma}  \notag\\
&+\psi^{\dagger}_{2n,\alpha} \phi^{\dagger}_{2n-1,\beta} \phi_{2n-1,\delta} \psi_{2n-1,\gamma}+h.c.\big) \ ,\label{eq:BFInteraction}
\end{align}
with $\boldsymbol{\mu}=(\alpha,\beta,\gamma,\delta)= (-1/2,0,1/2,-1)$.
The first term corresponds to Fig.~\ref{fig:3}a whereas the second term is shown in Fig.~\ref{fig:3}b.
According to \eqref{eq:coupling_CG}, the coupling constant for this specific scattering process is given by
\begin{align}
 g^{bf}_{\boldsymbol{\mu}}=\frac{\sqrt{2}}{3}\left(g_{bf,3/2}-g_{bf,1/2}\right) \ .
\end{align}
We emphasize that the spin exchange term does not change the staggered occupation of fermions such that the four-fermion term \eqref{eq:fourfermi} still does not contribute.
We anticipate that this applies to the elastic scattering terms as well.
Accordingly, the fermions are completely determined by their parity (even/odd sites) and we may therefore drop the spin label completely, such that 
\begin{align}
 \fh{\psi_{2n}\equiv\psi_{2n,-1/2} \quad , \quad \psi_{2n+1}\equiv\psi_{2n+1,1/2} \ .}
\end{align}
Employing the Schwinger boson representation and taking into account that the overlap integral $U^{bf}_{\mathbf{n}}$ does not depend on the specific lattice site $n$, the spin exchange Hamiltonian can be written as
\begin{align}
H^{bf}_{I_{se}} &= \frac{1}{2}U^{bf}_{\mathbf{n}}g^{bf}_{\boldsymbol{\mu}} \sum_n \left( \psi^{\dagger}_{n} L_{+,n} \psi_{n+1} + h.c.\right) \, .
\end{align}
The elastic scattering processes, on the other hand, are given by
\begin{align}
\label{eq:elastic_ham}
H^{bf}_{I_{el}} &= \frac{1}{2}\sum_{n',\beta} U^{bf}_{\mathbf{n}}g^{bf}_{\boldsymbol{\mu}}\psi^{\dagger}_{2n} \phi^{\dagger}_{2n,\beta} \phi_{2n,\beta} \psi_{2n}  \notag\\
                &+ \frac{1}{2}\sum_{n',\beta} U^{bf}_{\mathbf{n}}g^{bf}_{\boldsymbol{\mu}}\psi^{\dagger}_{2n} \phi^{\dagger}_{2n-1,\beta} \phi_{2n-1,\beta} \psi_{2n} \notag\\
                &+ \frac{1}{2}\sum_{n',\beta} U^{bf}_{\mathbf{n}}g^{bf}_{\boldsymbol{\mu}}\psi^{\dagger}_{2n+1} \phi^{\dagger}_{2n,\beta} \phi_{2n,\beta} \psi_{2n+1} \notag\\
                &+ \frac{1}{2}\sum_{n',\beta} U^{bf}_{\mathbf{n}}g^{bf}_{\boldsymbol{\mu}}\psi^{\dagger}_{2n+1} \phi^{\dagger}_{2n+1,\beta} \phi_{2n+1,\beta} \psi_{2n+1} \,, 
\end{align}
where $\beta\in\{ -1, 0\}$. 
We note that the coupling constants $g^{bf}_{\boldsymbol{\mu}}$ still depend on the magnetic substates and are, therefore, not identical for the different terms.
Moreover, we find that each $U^{bf}_{\mathbf{n}}$ is independent of $n$ in \eqref{eq:elastic_ham}, and identical in the first and second line (further denoted by $U^{bf}_{\mathbf{n}1}$) as well as in the third and fourth line (further denoted by $U^{bf}_{\mathbf{n}3}$), cf.\ Appendix~\ref{sec:app_overlap}.
For the first term in \eqref{eq:elastic_ham} with $\boldsymbol{\mu}=(-1/2,\beta,-1/2,\beta)$, we obtain
\begin{align}
\label{eq:elastic_ham_1}
H^{bf}_{I_{el,1}}&= \frac{g_{bf,1/2}+2g_{bf,3/2}}{6}\sum_{n'} U^{bf}_{\mathbf{n}1} \psi^{\dagger}_{2n} b^{\dagger}_{2n} b_{2n} \psi_{2n} \notag \\
                 &+\frac{g_{bf,3/2}}{2}\sum_{n'} U^{bf}_{\mathbf{n}1}\psi^{\dagger}_{2n} d^{\dagger}_{2n} d_{2n} \psi_{2n} \ , 
\end{align}
where we used \eqref{eq:coupling_CG} again.
The second term in \eqref{eq:elastic_ham} is the same as the first one upon replacing $b_{2n}\to d_{2n-1}$ and $d_{2n}\to b_{2n-1}$.
The third term in \eqref{eq:elastic_ham}, however, is different owing to $\boldsymbol{\mu}=(1/2,\beta,1/2,\beta)$ corresponding to the different fermionic parity and reads
\begin{align}
\label{eq:elastic_ham_3}
H^{bf}_{I_{el,3}}&=\frac{g_{bf,1/2}+2g_{bf,3/2}}{6}\sum_{n'}U^{bf}_{\mathbf{n}3}\psi^{\dagger}_{2n+1} b^{\dagger}_{2n} b_{2n} \psi_{2n+1} \notag \\
                 &+\frac{2g_{bf,1/2}+g_{bf,3/2}}{6}\sum_{n'}U^{bf}_{\mathbf{n}3}\psi^{\dagger}_{2n+1} d^{\dagger}_{2n} d_{2n} \psi_{2n+1} \ .
\end{align}
The fourth term in \eqref{eq:elastic_ham} is the same as the third one upon replacing $b_{2n}\rightarrow d_{2n+1}$ and $d_{2n}\rightarrow b_{2n+1}$.
Employing the Schwinger-boson representation $b^\dagger_{n}b_{n}=\ell+L_{z,n}$ and $d^\dagger_{n}d_{n}=\ell-L_{z,n}$, the first term \eqref{eq:elastic_ham_1} can be written as
\begin{align}
H^{bf}_{I_{el,1}}&=\frac{g_{bf,1/2}-g_{bf,3/2}}{6}\sum_{n'}U^{bf}_{\mathbf{n}1}\psi^{\dagger}_{2n}\psi_{2n}L_{z,2n} \notag \\
                 &+\frac{g_{bf,1/2}+5g_{bf,3/2}}{6}\ell\sum_{n'}U^{bf}_{\mathbf{n}1}\psi^{\dagger}_{2n}  \psi_{2n} \ .
\end{align}
Similarly, the third term \eqref{eq:elastic_ham_3} is given by
\begin{align}
H^{bf}_{I_{el,3}}&=\frac{g_{bf,3/2}-g_{bf,1/2}}{6}\sum_{n'}U^{bf}_{\mathbf{n}3}\psi^{\dagger}_{2n+1}\psi_{2n+1}L_{z,2n} \notag \\
                 &+\frac{g_{bf,1/2}+g_{bf,3/2}}{2}\ell\sum_{n'}U^{bf}_{\mathbf{n}3}\psi^{\dagger}_{2n+1}  \psi_{2n+1} \ ,
\end{align}
and similar expressions are also obtained for the second and fourth term by replacing $L_{z,2n}\to-L_{z,2n\mp1}$.
Accordingly, we obtain a contribution
\begin{align}
\label{eq:lzcont}
H^{bf}_{I_{el},L_z}&=\tilde{g}_{bf}\sum_{n'}U_{\mathbf{n}3}^{bf}\psi^\dagger_{2n+1}\psi_{2n+1}\left(L_{z,2n+1}-L_{z,2n}\right) \notag \\
                   &+\tilde{g}_{bf}\sum_{n'}U_{\mathbf{n}1}^{bf}\psi^\dagger_{2n}\psi_{2n}\left(L_{z,2n}-L_{z,2n-1}\right) \,
\end{align}
with $\tilde{g}_{bf}=(g_{bf,1/2}-g_{bf,3/2})/6$ that does not appear in the target Hamiltonian \eqref{eq:QL_Hamiltonian}.
The difference of the $L_z$ operators can be expressed in terms of the Gauss's law operator, $L_{z,n}-L_{z,n-1}=G_n+\psi^\dagger_n\psi_n + [(-1)^n-1]/2$. Any term proportional to $G_n$ exactly vanishes upon acting on physical states so that we can disregard them.
In summary, the elastic scattering terms result in bilinear, parity-dependent fermionic contributions that account to the potential energy of the fermions:
\begin{align}
 \label{eq:el_ferm}
 H^{bf}_{I_{el}}&=U_{\mathbf{n}3}^{bf}\Big((g_{bf,1/2}+g_{bf,3/2})\ell+\tilde{g}_{bf}\Big)\sum_{n'}\psi^\dagger_{2n+1}\psi_{2n+1} \notag \\ 
                &+U_{\mathbf{n}1}^{bf}\Big(\frac{g_{bf,1/2}+5g_{bf,3/2}}{3}\ell+\tilde{g}_{bf}\Big)\sum_{n'}\psi^\dagger_{2n}\psi_{2n} \, .
\end{align}

\subsection{Cold-atom QED Hamiltonian}

Summing up all contributions that originate from the kinetic, potential and interaction terms, the cold-atom QED Hamiltonian takes the form
\begin{align}
\label{eq:hamiltonian_cQED}
H &=  \frac{g_{b,0}-g_{b,2}}{6}U_{\mathbf{n}}^b\sum_n L^2_{z,n} + \Delta_{b,0} \sum_n (-1)^n L_{z,n}  \notag \\
& + \frac{1}{2}U^{bf}_{\mathbf{n}}g^{bf}_{\boldsymbol{\mu}}\sum_n   \left( \psi^{\dagger}_{n} L_{+,n} \psi_{n+1} + h.c.\right)\notag \\
&+ \sum_{n} \left( V^{f}_{n} \psi^{\dagger}_{n} \psi_{n} + V^{b}_{n} b^{\dagger}_{n} b_{n} + V^{d}_{n} d^{\dagger}_{n} d_{n} \right)\, .
\end{align}
The terms in the last line include the potential energy contributions, in particular the energy due to the trapping of the atoms as well as the bilinear term \eqref{eq:el_ferm}.
Again, we emphasize that the fermionic contribution $V^{f}_{n}$ depends on the parity of $n$. 
Introducing the energy difference $\Delta_f$ according to $V^f_{2n}=V^{f}_{0}-\Delta_f/2$ and $V^{f}_{2n+1}=V^{f}_0+\Delta_f/2$, the fermionic potential contribution is given by
\begin{align}
 H^{f}_{V}= V^{f}_0 \sum_n \psi^{\dagger}_n \psi_n -\frac{\Delta_f}{2}  \sum_n (-1)^n \psi^{\dagger}_n \psi_n \, .
\end{align}
Owing to total particle number conservation, the first term does not contribute to the dynamics and can thus be disregarded.
The bosonic potential term is treated in a similar fashion.
In fact, defining $\Delta_{b,1}$ according to $V^{b}_{n}=V^{b}_{0}+(-1)^n\Delta_{b,1}/2$ and $V^{d}_{n}=V^{b}_{0}-(-1)^n\Delta_{b,1}/2$ and using $L_{z,n}=(b^\dagger_n b_n-d^\dagger_n d_n)/2$, we obtain
\begin{align}
H^{b}_{V}= \Delta_{b,1}\sum_n (-1)^n L_{z,n} \, ,
\end{align}
where we disregarded an irrelevant constant proportional to $V^b_0$ which only depends on $\ell$.
Adding the second contribution in \eqref{eq:hamiltonian_cQED} and defining $\Delta_b\equiv\Delta_{b,0}+\Delta_{b,1}$, we obtain
\begin{align}
 H^{b}_{V}+\Delta_{b,0} \sum_n (-1)^n L_{z,n} = \Delta_{b} \sum_n (-1)^n L_{z,n} \ .
\end{align}
Comparison of the cold-atom QED Hamiltonian with the target Hamiltonian \eqref{eq:QL_Hamiltonian} then shows that we have a term linear in $L_{z,n}$.
However, this term does not contribute. To see this, we transform the Hamiltonian into the interaction picture.
To this end, we split the cold-atom QED Hamiltonian into two parts, $H=H_0+H_1$, with 
\begin{subequations}
\begin{align}
 \!H_0&=\Delta_{b} \sum_n(-1)^n \Big(L_{z,n}-\frac{1}{2}\psi^{\dagger}_n \psi_n\Big) \ , \\
 \!H_1&=\chi_{BB}\sum_n L^2_{z,n} + \Delta \sum_n (-1)^n \psi^{\dagger}_n \psi_n \notag \\
   \! &+\frac{\chi_{BF}}{2}\sum_n \left( \psi^{\dagger}_{n} L_{+,n} \psi_{n+1} + h.c.\right) \ .
\end{align}
\end{subequations}
Here, we introduced the detuning $2\Delta \equiv \Delta_b-\Delta_f$, the effective boson self-interaction $\chi_{BB}$ and the boson-fermion interaction $\chi_{BF}$,
\begin{subequations}
\begin{align}
\chi_{BB} &= \frac{g_{b,0}-g_{b,2}}{6}U_{\mathbf{n}}^b \, ,\\
\chi_{BF} &= \frac{\sqrt{2}(g_{bf,3/2}-g_{bf,1/2})}{3}U_{\mathbf{n}}^{bf} \,  .
\end{align}
\end{subequations}
The index $\mathbf{n}$ is the same as in \eqref{eq:BosonInteraction} and \eqref{eq:BFInteraction} and the overlaps $U_\mathbf{n}$ are determined in \eqref{eq:Overlaps} of the appendix.
Upon acting with the unitary transformation $U(t)=\exp(-iH_0t)$, it is straightforward to show that $H'_1=U^\dagger(t)H_1U(t)$.
Performing the canonical transformation $\psi_n\to(-i)^n\psi_n$, we can finally identify $H'_1$ with the quantum link Hamiltonian \eqref{eq:QL_Hamiltonian}
\begin{align}
 \label{eq:Hamiltonian}
 H_{\text{QL}}&=\frac{g^2a_S}{2}\sum_n L^2_{z,n} + M \sum_n (-1)^n \psi^{\dagger}_n \psi_n \notag \\
                   &-\frac{i}{2a_S\sqrt{\ell(\ell+1)}}\sum_n \left( \psi^{\dagger}_{n} L_{+,n} \psi_{n+1} - h.c.\right) \ 
\end{align}
where we introduced the abbreviations
\begin{subequations}
\begin{align}
\frac{\chi_{BB}}{\Delta}&\equiv \frac{g^2a_S}{2M} \ , \\
\frac{\chi_{BF}}{\Delta}&\equiv \frac{1}{a_S M \sqrt{\ell(\ell+1)}} \ ,
\end{align}
\end{subequations}
and time is measured in units of $M$ instead of $\Delta$.
We note that we have to take the limit $\ell\to\infty$ in order to recover the Hamiltonian formulation of lattice QED corresponding to \eqref{eq:KS_Hamiltonian}.

\section{Microscopic parameters \label{sec:IV}}

At this point, we are now able to determine the accessible parameters for an experimental implementation of the Schwinger model via a mixture of bosonic $^{23}\rm{Na}$ and fermionic $^{6}\rm{Li}$ atoms \cite{Scelle2013a}, which is determined by the parameters $\chi_{BB}$, $\chi_{BF}$, $\Delta$ and the occupation numbers of the links.

The wave functions that appear in the overlap integrals for $\chi_{BF}$ and $\chi_{BB}$ are determined  by the lattice spacing $a$, the lattice depths $V^b$, $V^f_{1}$ and $V^f_{2}$ appearing in \eqref{eq:LatticePotential}, the difference of scattering lengths that drives the spin-changing collisions as well as the atomic density on each site.
For Na + Na in the $f_b=1$ manifold, one obtains $a_{b,0}-a_{b,2} = 5 \, a_0 $ \cite{Stamper-Kurn2013}, where $a_0$ is the Bohr radius. 
For the Na + Li collisions, we have $a_{bf,3/2}-a_{bf,1/2} = 0.9 \, a_0$ \cite{Note1}.
Further, we choose $a = \SI{4}{\micro \meter}$, $V_1^b = 20 \, E_R $, $V^f_{1} = 2.76 \, E_R $ and $V^f_{2} =-1.85 \, E_R$ with the recoil energy  $ E_R = \frac{\hbar^2 }{2M_b} \left( \frac{2\pi}{a}\right)^2 \,$. Here the number of lattice sites can range from a few sites up to $N \lesssim 100$, where the limit arises typically due to different experimental constraints. 
A judicious choice of the orthogonal confinement allows us to use the same wave function for bosons and fermions in the orthogonal direction $\varphi_{s}(y) = \varphi_{s}(z)$ with $\omega_{\perp}/2\pi = \SI{5}{\kilo \hertz}$.
This potential results in $\chi_{BF}/h =\SI{0.05}{\hertz}$, $\chi_{BB}/h = \SI{0.58}{\hertz}$. We choose the occupation of the bosonic links to be $N_B \approx100$ atoms per well, such that the estimated time scale for three-body loss is of the order of several seconds. This means that the Bose-Fermi coupling will be typically of the order of $\chi_{BF}N_B/h\approx \SI{5}{\hertz}$. It will lead to a hopping of fermions between neighboring lattice sites if the detuning is not too large and we therefore choose a detuning of $\Delta/h \approx \SI{10}{\hertz}$. Further, the chosen experimental parameters ensure that the bosonic and fermionic tunneling $J_B/h \approx \SI{0.25}{\hertz}$, $J_F/h \approx \SI{0.25}{\hertz}$ (see appendix  \eqref{eq:OverlapIntegral}) is ten times slower than the expected  gauge field dynamics.  
So we can indeed neglect the direct tunneling for a description of the cold-atom system as done in the previous derivation.

Given the microscopic parameters of the model, we have in mind applications to strong-field phenomena such as Schwinger pair production or string breaking, for which we aim to perform benchmark simulations of the cold-atom Hamiltonian (\ref{eq:hamiltonian_cQED}).
To study Schwinger pair production, the electric field $E$ is supposed to exceed the critical field strength $E_c=M^2/g$ such that the amplitude ${E} / {E_c} \gtrsim 1$.
In the atomic system, this is given by
\begin{align}
 \frac{g^2 L_{z}}{M^2} = \frac{ \sqrt{\ell(\ell+1)} (N_b - N_d)\chi_{BF} \chi_{BB}}{\Delta^2} \label{Eq:alpha} 
\end{align}
with $N_b=\langle b_n^\dagger b_n\rangle$ and $N_d=\langle d_n^\dagger d_n\rangle$. Further, we approximate $\sqrt{\ell(\ell+1)} \approx N_B/2$ and $N_b - N_d \approx N_B$.
The achievable electric field exceeds the critical field strength with $E/E_c \approx 1.5$ for the given experimental parameters.

This puts the exciting prospect of studying Schwinger pair production using ultracold atoms within reach of current experimental techniques. 
For a successful quantum simulation of Schwinger pair production, however, it does not suffice to implement only the appropriate Hamiltonian.
Additionally, we also need to verify (i) that a proper initial state can be prepared, (ii) that the QED dynamics is indeed accessible to the experimental protocol, and (iii) that we can read-out the relevant quantities experimentally.  
In fact, all these requirement can be fulfilled with current experimental techniques for a range of important strong-field phenomena, as further described in section~\ref{sec:VIB}.

\section{Functional Integral approach \label{sec:V}}
In this section we outline the functional integral (FI) method 
to investigate the real-time dynamics of fermions coupled to bosonic fields
and for convenience we perform the following theoretical calculations in natural units. Since identical fermions cannot occupy the
same state, their quantum nature is highly relevant and a consistent quantum
theory of nonequilibrium Schwinger pair production including backreactions
is envisaged. Based on the defining functional integral of the quantum
theory, a systematic expansion can be achieved where the corrections are
given in terms of a small (dimensionless coupling) parameter for
strong-field phenomena as reviewed in Ref.~\cite{2014PhRvD..90b5016K}. At
lowest order in this expansion one recovers a classical-statistical field
theory for bosonic fields or the so-called 
Truncated Wigner approximation. The inclusion of fermions requires to go
beyond lowest order, which we describe below and
apply to the cold-atom Hamiltonian \eqref{eq:Hamiltonian} in section
\ref{sec:VI}. To study the strong-field regime of QED, the field strength
needs to be of the order of the critical field $E_c=M^2/g$.
The corresponding cold atom setup is characterized by $E_c=g|N_b - N_d|/2
\sim M^2/g$, where $N_b,N_d$ denote the number of atoms in the Bose-Einstein
condensates. 
For $N_b,N_d \sim \mathcal{O}(\ell)\gg 1$, the FI approach of
Refs.~\cite{1998NuPhB.511..451A,2013PhRvD..87j5006H,2014PhRvD..90b5016K, 2011PhRvL.107f1301B,2014PhRvD..90d5034H} allows us to study the dynamics in this regime.

To make contact with the Truncated Wigner approach, we start from 
a purely bosonic theory \cite{2010AnPhy.325.1790P}.
The expectation value of an observable $O$ is given by the trace
\begin{align}
 \label{eq:observable}
 \braket{O(t)} \equiv \Tr \left\{ O\rho(t) \right\}\ ,
\end{align}
where the time-dependent density operator $\rho(t)$ in the Schr\"odinger picture obeys the von-Neumann equation
\begin{align}
i \partial_t \rho(t) = [H,\rho(t)]  \ .
\end{align}
Its discretized version 
\begin{align}
\label{eq:evolution_rho}
\rho(t+\Delta t) =   \rho(t) -  i\Delta t [ H \rho(t) - \rho(t) H ] 
\end{align}
may be used as a starting point for deriving a
functional integral expression of the time evolution 
for the density matrix. 
We perform a Wigner transformation of the discretized 
von-Neumann equation  \eqref{eq:evolution_rho}
\begin{align}
 &\rho_W(\varphi_1;t_1)  = \rho_W(\varphi_1;t_0) \notag \\
 &\qquad - i\Delta t [ (H \rho)_W(\varphi_1;t_0) - (\rho H)_W(\varphi_1;t_0)] \ ,
\end{align}
where the transformation only acts on bosonic degrees of freedom in the density matrix and the Hamiltonian.
The Wigner transform, which depends on the center field $\varphi_1$, is denoted by $(\,\cdot\,)_W$, and we refer to appendix \ref{sec:app_wigner} for further details.
Here, $\varphi_1$ may carry additional indices that will be suppressed in the following.
We emphasize that there are no operators present anymore as they are replaced by c-number variables.
Employing the product formula for the Wigner transform \eqref{eq:app_wignerprod}, the last equation can be written as
\begin{align}
 &\rho_W(\varphi_1;t_1) =  \int \frac{d^2 \varphi_0}{\pi}\int \frac{d^2\eta_1}{\pi}  \rho_W(\varphi_0;t_0) \notag \\
 &\!\times \exp \left\{ \eta^{\ast}_1(\varphi_0 -  \varphi_1) - \eta_1(\varphi_0^* -  \varphi_1^*)\right\} \notag \\
 &\!\times \exp \Big\{ - i \Delta t \Big[H_W(\varphi_1 + \tfrac{1}{2}\eta_1)   - H_W(\varphi_1 - \tfrac{1}{2}\eta_1)\Big] \Big\} \,. 
\end{align}
The product formula of Wigner transforms introduces an additional integration with respect to the difference field $\eta_1$.
The naming 'center field' and 'difference field' is motivated by the Schwinger-Keldysh formalism as reviewed in Ref.~\citep{2014PhRvD..90b5016K}, in which the fields on the forward and backward branch of the closed-time path correspond to $\varphi^{\pm}=\varphi\pm\eta/2$.
Iterating this expression, we obtain the functional integral representation 
\begin{align}
 \label{eq:boson_single}
  &\rho_W(\varphi_N;t_N) = \prod_{k=0}^{N-1} \int \frac{d^2 \varphi_k}{\pi} \prod_{k=1}^{N}\int \frac{d^2\eta_k}{\pi}  \rho_W(\varphi_0;t_0) \notag \\
 &\!\times \exp\Big\{\sum_{k=1}^N \eta^{\ast}_k(\varphi_{k-1} -  \varphi_k) - \eta_k(\varphi_{k-1}^* -  \varphi_k^*)\Big\} \notag \\
 &\!\times \exp\Big\{- i \Delta t\sum_{k=1}^N \left[ H_W(\varphi_k \!+\! \tfrac{1}{2}\eta_k)   \!-\!  H_W(\varphi_k \!-\! \tfrac{1}{2}\eta_k)\right] \!\Big\} 
\end{align}
with $t_N=t_0+N\Delta t$. 
The exponent in the last expression can be written as
\begin{align}
i S[\varphi,\eta] &\equiv i 
\Delta t \sum_{k=1}^N \left[i\eta^{\ast}_k \frac{\varphi_{k} -  \varphi_{k-1}}{\Delta t} - H_W(\varphi_k \!+\! \tfrac{1}{2}\eta_k)\right] \notag \\
&- i \Delta t\sum_{k=1}^N \left[i \eta_k  \frac{\varphi_{k}^* -  \varphi_{k-1}^*}{\Delta t}  \!-\!  H_W(\varphi_k \!-\! \tfrac{1}{2}\eta_k)\right] \,,   
\end{align}
which is  a discretized version of the Schwinger-Keldysh action. 
Expanding the Hamiltonian up to linear order in the difference field $\eta_k$,
\begin{align}
 &H_W(\varphi_k \pm \tfrac{1}{2} \eta_k) = \notag \\
 &H_W(\varphi_k) \pm \frac{\partial H_W(\varphi_k)}{\partial \varphi_k} \frac{\eta_k}{2} \pm \frac{\partial H_W(\varphi_k)}{\partial \varphi^{\ast}_k} \frac{\eta^{\ast}_k}{2} +\mathcal{O}(\eta_k^2) \, ,
\end{align}
we obtain
\begin{align}
 &\rho_W(\varphi_N;t_N) = \prod_{k=0}^{N-1} \int \frac{d^2 \varphi_k}{\pi} \prod_{k=1}^{N}\int \frac{d^2\eta_k}{\pi}  \rho_W(\varphi_0;t_0) \notag \\
 &\!\times \exp\Big\{\sum_{k=1}^N \eta^{\ast}_k(\varphi_{k-1} -  \varphi_k) - \eta_k(\varphi_{k-1}^* -  \varphi_k^*)\Big\} \notag \\
 &\!\times \exp\Big\{ -i \Delta t \sum_{k=1}^N \Big[ \frac{\partial H_W(\varphi_k)}{\partial \varphi_k}\eta_k + \frac{\partial H_W(\varphi_k)}{\partial \varphi^{\ast}_k} \eta^{\ast}_k\Big] \Big\} \,. \label{eq:TWABosonsFI}
\end{align}
Within this approximation, the difference field $\eta_k$ can be integrated out as it appears only linearly in the exponent.
This gives
\begin{align}
\label{eq:bosonic_wigner}
&\rho_W(\varphi_N;t_N) = \prod_{k=0}^{N-1} \int \frac{d^2 \varphi_k}{\pi} \rho_W(\varphi_0;t_0) \prod_{k=1}^N \delta(f(\varphi_k,\varphi_{k-1})) \, .
\end{align}
The arguments of the complex Dirac-delta functions
\begin{align}
 f(\varphi_k,\varphi_{k-1})=\varphi_{k-1} - \varphi_k -i \Delta t  \frac{\partial H_W(\varphi_k)}{\partial \varphi^{\ast}_k}  
\end{align}
impose the discrete time evolution equation \cite{2010AnPhy.325.1790P}
\begin{align}
 \label{eq:bosonic_eom}
 i\frac{\varphi_k - \varphi_{k-1}}{\Delta t} =  \frac{\partial H_W(\varphi_k)}{\partial \varphi^{\ast}_k} \, .
\end{align}
In fact, this equation can be solved implicitly, such that $\varphi_{k}=g(\varphi_{k-1})$.
To actually integrate out $\varphi_k$ for $k\in\{0,\ldots,N-1\}$, we have to change the argument of the delta function, such that
\begin{align}
 \delta\left(f(\varphi_k,\varphi_{k-1})\right)=\left| \frac{ \partial ( f , f^*)}{\partial (\varphi_k, \varphi_k^*) }\right|^{-1}  \delta  \left( \varphi_k - g(\varphi_{k-1}) \right) \, . \label{eq:EOMBosons}
\end{align}
A similar derivation can be performed for real scalar field theories
\cite{2007PhRvA..76c3604B}.
The Jacobi determinant takes the form
\begin{align}
 \left| \frac{ \partial ( f , f^*)}{\partial (\varphi_k, \varphi_k^*) }\right|=\left|1+i\Delta t K\right|=1+\mathcal{O}(\Delta t^2) \ ,
\end{align}
where we used
\begin{align}
 \det(1+\epsilon K)=1+\epsilon \Tr{K}+\mathcal{O}(\epsilon^2)
\end{align}
with
\begin{align}
 K = \begin{pmatrix} \displaystyle
 \frac{\partial^2 H_W(\varphi_k)}{\partial \varphi^{\ast}_k \partial \varphi_k} & \displaystyle
 \frac{\partial^2 H_W(\varphi_k)}{\partial \varphi^{\ast}_k \partial \varphi^{\ast}_k} \\ \displaystyle
 - \frac{\partial^2 H_W(\varphi_k)}{\partial \varphi_k \partial \varphi_k} & \displaystyle
 - \frac{\partial^2 H_W(\varphi_k)}{\partial \varphi^{\ast}_k \partial \varphi_k} 
 \end{pmatrix} 
\end{align}
and $\epsilon$  being a small number.
This  shows that the Jacobi determinant does not contribute \cite{2005PhRvC..72a4907J}, since it affects the dynamics only at order $\mathcal{O}(\Delta t^2)$.
The Wigner function $\rho_W(\varphi_N;t_N)$ for arbitrary times $t_N$ is obtained by sampling over initial conditions $\varphi_0$, which are then evolved by \eqref{eq:bosonic_eom}. 
Expectation values of bosonic observables $O$ are then given by
\begin{equation}
 \braket{O(t)} =\int{\frac{d^2\varphi_N}{\pi} \rho_W(\varphi_N;t_N)O_W(\varphi_N)} \ .
\end{equation}

\subsection{Interacting boson-fermion theory \label{ssec:BosonsFermions}}

We now include the quantum corrections induced by the coupling to the fermions.
Here, Wigner transforms are only calculated with respect to the bosonic variables whereas the fermionic operators and their appropriate time-ordering still has to be taken into account.
For the sake of simplicity, we denote the fermionic fields by $\psi$ and emphasize that they carry additional indices which will be suppressed in the following.
Furthermore, we assume that the Hamiltonian $H = H_B + H_F$ can be separated into a purely bosonic part $H_B$ and a quadratic fermionic part, $H_F = \psi^{\dagger} h_F \psi $, where $h_F$ is a matrix and may contain bosonic degrees of freedom.

The Wigner transform of the discretized time evolution equation \eqref{eq:evolution_rho} for a single time step is now given by
\begin{align}
 &\rho_W(\psi,\varphi_1;t_1) =\int \frac{d^2 \varphi_0}{\pi} \int \frac{d^2\eta_1}{\pi}  \notag \\
 &\!\times \exp \left\{ \eta^{\ast}_1(\varphi_0 -  \varphi_1) - \eta_1(\varphi_0^* -  \varphi_1^*)\right\} \notag \\
 &\!\times \exp \Big\{ - i \Delta t H_W(\psi,\varphi_1 + \tfrac{1}{2}\eta_1)\Big\} \rho_W(\psi,\varphi_0;t_0) \notag \\
 &\!\times \exp \Big\{  i \Delta t H_W(\psi,\varphi_1 - \tfrac{1}{2}\eta_1) \Big\} \, ,
\end{align}
where we emphasize again that the Wigner transform $H_W(\psi,\varphi)$ depends on both fermionic operators $\psi$ and c-number variables $\varphi$.  
Iterating this expression, we obtain again a functional integral representation of the time evolution of the Wigner function
\begin{align}
 &\rho_W(\psi,\varphi_N;t_N) = \prod_{k=0}^{N-1} \int \frac{d^2 \varphi_k}{\pi} \prod_{k=1}^{N}\int \frac{d^2\eta_k}{\pi} \notag \\ 
 &\! \times \exp\Big\{\sum_{k=1}^N \eta^{\ast}_k(\varphi_{k-1} -  \varphi_k) - \eta_k(\varphi_{k-1}^* -  \varphi_k^*)\Big\} \notag \\
 &\! \times T \exp\Big\{ - i \Delta t \sum_{k=1}^N H_W(\psi,\varphi_k + \tfrac{1}{2}\eta_k) \Big\} \rho_W(\psi,\varphi_0;t_0) \notag \\ 
 &\!\times \bar{T}\exp \Big\{i \Delta t \sum_{k=1}^N H_W(\psi,\varphi_k - \tfrac{1}{2} \eta_k) \Big\} \, ,
\end{align}
where we introduced the time ordering operator $T$ and the anti-time ordering operator $\bar{T}$.
Based on the assumption that $H = H_B + H_F$, we may also separate its Wigner transform into a purely bosonic and a fermionic contribution
\begin{align}
 H_W(\psi,\varphi_k) &= H_{BW}(\varphi_k) + H_{FW}(\psi,\varphi_k) \, .
\end{align}
Furthermore, we assume that the initial density matrix factorizes into a purely bosonic part $\rho_B$ as well as a purely fermionic contribution $\rho_F$. 
Accordingly, the Wigner transform at initial times only affects the bosonic part, such that
\begin{align}
 \rho_W(\psi,\varphi_0;t_0)=\rho_{BW}(\varphi_0;t_0)\rho_F(\psi;t_0) \ .
\end{align}
also factorizes. However, the factorization property of the density matrix may be lost during the time evolution.
We integrate out the fermions to get the time evolution of bosonic observables $O_B$.
Denoting the trace over fermionic operators by $\Tr_F$, we obtain
\begin{align}
 \label{eq:fermionic_fi}
 \!\!\!&\operatorname{Tr}_F \!\rho_W(\psi,\varphi_N;t_N) \!=\! \prod_{k=0}^{N-1}\! \int \! \frac{d^2 \varphi_k}{\pi} \prod_{k=1}^{N}\! \int \! \frac{d^2\eta_k}{\pi}\rho_{BW}(\varphi_0;t_0) \notag \\
  \!\!\!&\!\!\times \exp\!\Big\{\sum_{k=1}^N \eta^{\ast}_k(\varphi_{k-1} -  \varphi_k) - \eta_k(\varphi_{k-1}^* -  \varphi_k^*)\Big\} \notag \\
  \!\!\!&\!\!\times \exp\!\Big\{\!\!- \!\!i \Delta t \! \sum_{k=1}^N \! \left[ H_{BW}(\varphi_k \!+\! \tfrac{1}{2}\eta_k)   \!-\!  H_{BW}(\varphi_k \!-\! \tfrac{1}{2}\eta_k)\right] \!\!\Big\} \notag \\
 \!\!\! &\!\!\times  \operatorname{Tr}_F\left[ T \exp\Big\{ - i \Delta t \sum_{k=1}^N  H_{FW}(\psi,\varphi_k + \tfrac{1}{2}\eta_k) \Big\} \right. \notag \\
 \!\!\! &\!\!\times \left. \rho_F(\psi,t_0) \bar{T}\exp \Big\{i \Delta t \sum_{k=1}^N H_{FW}(\psi,\varphi_k - \tfrac{1}{2}\eta_k)  \Big\}\right] \, .
\end{align}
Owing to the fact that the purely bosonic part is the
same as in \eqref{eq:boson_single}, we focus on the fermionic contributions in the following.
Since $H_F$ is taken to be quadratic in the fermionic operators, it is convenient to 
 introduce the abbreviation
\begin{align} 
 \psi^{\dagger}h_{FW}(\varphi) \psi &\equiv H_{FW}(\psi,\varphi) \ . \label{eq:HamFermWeyl1}  
\end{align} 
Here, $h_{FW}(\varphi)$ is a matrix that only depends on bosonic variables but not on fermionic operators.
Further, we introduce  $\varphi^\pm_k$ as the linear combination 
of center field $\varphi_k$ and difference field $\eta_k$, 
\begin{align}
\varphi^\pm_k = \varphi_k \pm \tfrac{1}{2} \eta_k \ ,
\end{align}
and we assume that the initial fermionic density matrix can be written as
\begin{align}
\rho_F = \mathcal{Z}^{-1} \exp[\psi^{\dagger} h_{FW}(\varphi_0) \psi ] \ ,
\end{align}
where $\mathcal{Z}$ is an appropriate normalization. 
By utilizing the identity \cite{PhysRevLett.110.137002}
\begin{align}
 \Tr_F{e^{\psi^\dagger M_1\psi}\ldots e^{\psi^\dagger M_n\psi}}=\det(1+e^{M_1}\ldots e^{M_n}) 
\end{align}
with matrices $M_k$ for $k = 1, \ldots, n$, we may explicitly perform the fermionic trace in \eqref{eq:fermionic_fi}.
Introducing the evolution matrix
\begin{align}
 S_{k,m}(\varphi) \equiv e^{-i \Delta t\, h_{FW}(\varphi_k) } \ldots e^{-i \Delta t\, h_{FW}(\varphi_m)  } \, ,
\end{align}
where $S_{k,m}(\varphi)$ depends on the string of fields $\varphi_k, ..., \varphi_m$ for $k\geq m $, we obtain 
\begin{align}
\label{eq:fermionic_trlog}
\operatorname{Tr}_F &\left[ T \exp\{\ldots\}\rho_F(\psi;t_0) \bar{T}\exp\{\ldots\}\right]= \notag \\
&\mathcal{Z}^{-1} \exp \operatorname{Tr} \log \big(1 + S_{N,1}(\varphi^{+})  e^{ h_{FW}(\varphi_0) } S^{\dagger}_{N,1}(\varphi^{-})\big) \, 
\end{align}
for the last two lines of \eqref{eq:fermionic_fi}.
For later convenience, we also define the fermionic  propagator
\begin{align}
 D_k\equiv D(t + k \Delta t ) =   S_{k,1}(\varphi)(1+e^{-h_{FW}(\varphi_0)})^{-1} S^{\dagger}_{k,1}(\varphi) \ ,
\end{align}
whose time evolution at order $\mathcal{O}(\Delta t)$ is governed by
\begin{align}
 &D_{k+1} - D_k = i \Delta t \left[ D_k , h_{FW}(\varphi_{k+1}) \right] \ . \label{eq:EOMofD} 
\end{align}
The components of $D_k$ can be identified with the equal-time correlation function $\braket{\psi^\dagger_m\psi_n}$ as will be shown below. 
This discrete equation can be solved via a mode expansion of the operator $\psi_m$ as  illustrated in Ref. \cite{2014PhRvD..90b5016K}. Knowing
the mode functions then allows one to compute fermionic correlation functions.

The expansion of the $\operatorname{Tr}\log$ in \eqref{eq:fermionic_trlog} up to linear order in the difference field $\eta_k$ then yields 
\begin{align}
 \label{eq:fermionic_fi2}
\! &\operatorname{Tr}_F\! \rho_W(\psi,\varphi_N;t_N)\! = \!\prod_{k=0}^{N-1}\! \int \frac{d^2 \varphi_k}{\pi} \prod_{k=1}^{N}\!\int \frac{d^2\eta_k}{\pi}\rho_{BW}(\varphi_0;t_0) \notag \\
\! &\!\times \! \exp\Big\{\sum_{k=1}^N \eta^{\ast}_k(\varphi_{k-1} -  \varphi_k) - \eta_k(\varphi_{k-1}^* -  \varphi_k^*)\Big\} \notag \\
\! &\!\times\! \exp\!\Big\{ \!\!-i \Delta t \sum_{k=1}^N \Big[ \frac{\partial H_{BW}(\varphi_k)}{\partial \varphi_k}\!+\Tr\Big({\frac{\partial h_{FW}(\varphi_k)}{\partial \varphi_k}  D_k}\Big) \Big]\eta_k \Big\}  \notag \\
\! &\!\times\! \exp\!\Big\{ \!\!-i \Delta t \sum_{k=1}^N \Big[ \frac{\partial H_{BW}(\varphi_k)}{\partial \varphi^{\ast}}\! + \Tr \Big(\frac{\partial h_{FW}(\varphi_k)}{\partial \varphi_k^*}D_k\Big) \Big]\eta_k^* \Big\}  .
\end{align}
Since the difference field $\eta_k$ appears only linearly in the exponent, we may integrate it out again.
As compared to the purely bosonic case, cf.~\eqref{eq:bosonic_eom}, the resulting complex Dirac-delta function now impose the discrete time evolution equation \cite{1999NuPhB.555..355A}
\begin{align}
i\frac{\varphi_k - \varphi_{k-1}}{\Delta t}
 =  \frac{\partial H_W(\varphi_k)}{\partial \varphi^{\ast}_k} +\operatorname{Tr} \left(  \frac{\partial h_{FW}(\varphi_k)}{\partial \varphi^{\ast}_k}  D_k  \right) \, , \label{eq:EOMwithCurrent}
\end{align}
which is implicitly solved via $\varphi_{k}=\tilde{g}(\varphi_{k-1})$.
As in the purely bosonic case, the Jacobi determinant does not contribute at leading order.
Accordingly, the Wigner function $\rho_W(\varphi_N;t_N)$ is obtained by sampling over initial conditions and subsequent time evolution of $\varphi$ and $D$ according to \eqref{eq:EOMwithCurrent} and \eqref{eq:EOMofD}, respectively.

Finally, to calculate the expectation value of bilinear fermionic observables $O_F=\psi^\dagger A\psi$ with $A$ being a matrix, we consider
\begin{align}
\braket{O_F(t)} &= \Tr \{ \psi^{\dagger} A \psi\, \rho(t) \} \notag \\
 & =  \int \frac{d^2 \varphi_N}{\pi} \Tr_F \left\{ \psi^{\dagger} A \psi\, \rho_W(\psi,\varphi_N;t)\right\} \ .
\end{align}
While the bosonic contributions are identical to \eqref{eq:fermionic_fi}, the fermionic trace now takes the form
\begin{align}
 \operatorname{Tr}_F &\left[ \psi^\dagger A \psi\, T \exp\Big\{ - i \Delta t \sum_{k=1}^N  H_{FW}(\psi,\varphi_k + \tfrac{1}{2}\eta_k) \Big\} \right. \notag \\
 & \times\left. \rho_F(\psi;t_0) \bar{T}\exp \Big\{i \Delta t \sum_{k=1}^N H_{FW}(\psi,\varphi_k - \tfrac{1}{2}\eta_k)  \Big\}\right] \, .
\end{align}
Utilizing the identity 
\begin{align}
 &\Tr_F{\psi^\dagger A\psi\ e^{\psi^\dagger M_1\psi}\ldots e^{\psi^\dagger M_n\psi}}= \notag \\
 &\det(1+e^{M_1}\ldots e^{M_n})\Tr\{(1+e^{-M_n}\ldots e^{-M_1})^{-1}A\} \ ,
\end{align}
we may again perform the fermionic trace explicitly.
Expanding the exponent up to linear order in the response field $\eta_k$ while setting it to zero otherwise, we finally obtain
\begin{align}
 \braket{O_F(t_N)}&=\prod_{k=0}^N \int \frac{d^2 \varphi_k}{\pi}\rho_{BW}(\varphi_0;t_0) \notag \\
                  &\times\prod_{k=1}^{N}\delta(\varphi_k - \tilde{g}(\varphi_{k-1}))\Tr \{D_N A \} \,.
\end{align}
Accordingly, the discrete time evolution of $\braket{O_F}$ is determined by
\begin{align}
 \braket{O_F(t_{k+1})}  - \braket{O_F(t_k)} = i\Delta t   \Tr \{ \left[D_k ,  h_{FW}(\varphi_{k+1}) \right] A \}   \,.
\end{align}
By explicitly introducing spatial indices and choosing $A_{mn}=\delta_{mm_1}\delta_{nn_1}$, we recover the evolution equation of $D_k$ as given in \eqref{eq:EOMofD}.
This shows that $D_k$ can indeed be identified with the equal-time correlation function $\braket{\psi^\dagger_m\psi_n}$.

\subsection{Equations of motion for the cold atom system\label{sec:ClasStatEom}}

To derive the equations of motion for the cold atom system, we apply the method from the previous section to the Hamiltonian \eqref{eq:Hamiltonian}.
To this end, we rewrite $H_{QL}$ in terms of the Schwinger boson operators $b_n$ and $d_n$.
Upon employing symmetric operator ordering and skipping an irrelevant constant, the Wigner transform of the Hamiltonian is given by \cite{UBHD-64217292} 
\begin{align}
H_{QL,W} &= \frac{g^2a_S}{4} \sum_n (|b_{n}|^4+|d_{n}|^4) + M \sum_n (-1)^n  \psi^{\dagger}_n \psi_n \notag \\ 
&- \frac{i}{2a_S\sqrt{\ell(\ell+1)}}\sum_n (\psi^{\dagger}_n b^{\ast}_n d_n \psi_{n+1} -h.c.) \ , 
\end{align}
where $b_n$ and $d_n$ are now c-numbers, but the $\psi_n$ and $\psi^{\dagger}_n$ are still fermionic operators.
In the following, time is treated as a continuous variable.
Accordingly, all equations correspond to their discrete versions up to order $\mathcal{O}(\Delta t^2)$.
The equations of motion for the fermionic correlator $D(t)$ are given by 
\begin{align}
i \partial_t D_{m,n}(t) = h_{FW,mk}(t) D_{k,n}(t) - D_{m,k}(t) h_{FW,kn}(t) \,, \label{eq:eom1}
\end{align}
where the fermionic matrix $h_{FW}(t)$ reads
\begin{align}
h_{FW,mn} = &-\frac{i}{2a_S \sqrt{\ell(\ell+1)}} b^{\ast}_m d_m \delta_{m+1,n} + M (-1)^m \delta_{m,n} \notag \\
&+\frac{i}{2a_S \sqrt{\ell(\ell+1)}}d^{\ast}_{m-1} b_{m-1} \delta_{m-1,n} \, .
\end{align}
The fermionic two-point function $D_{mn}=\langle\psi_m^\dagger\psi_n\rangle=\frac{1}{2}(\delta_{mn}-F_{nm})$ can be 
expressed in terms of the statistical two-point function $F_{mn}=\langle[\psi_m,\psi_n^\dagger]\rangle$.
Accordingly, the equations of motion for the bosonic degrees of freedom are given by
\begin{subequations}
\label{eq:eom3}
\begin{align}
i\partial_t b_{n} &= \frac{g^2a_S}{2} b^{*}_n b_n b_n  +\frac{i d_n}{4a_S \sqrt{\ell(\ell+1)}} F_{n+1,n} ,\\
i \partial_t d_{n} &= \frac{g^2a_S}{2} d^{*}_n d_n d_n -\frac{i b_n}{4a_S \sqrt{\ell(\ell+1)}} F_{n,n+1}   \, . 
\end{align}
\end{subequations}
We  specify initial conditions to solve the system of time evolution equations \eqref{eq:eom1} and \eqref{eq:eom3}.
The method outlined above allows us to use Gaussian initial states for the fermions. 
In the following, we focus on the vacuum of the  Hamiltonian 
\begin{align}
 H_{f,0}&=-\frac{i}{2a_S} \sum_{n} ( \psi_n^{\dagger}  \psi_{n+1} - h.c.)  +  M \sum_n(-1)^n \psi^{\dagger}_n \psi_n \ .
 \label{eq:ferm_free}
\end{align}
Since $H_{f,0}$ is quadratic, the ground state and the dispersion relation can be determined analytically, cf.\ Appendix~\ref{sec:app_particle}.
Given  an optical lattice with $N$ elementary cells, i.e.\ $2N$ lattice sites, the dispersion relation is given by two bands $\pm\omega_q$ with
\begin{align} 
 \omega_q=\sqrt{M^2+\pi_q^2} \quad , \quad \pi_q=\frac{1}{a_S}\sin\Big(\frac{\pi q}{N}\Big) \ , 
\end{align}
with $q\in\{0,\ldots,N-1\}$.
The corresponding mode function expansion of the fermionic field operator reads
\begin{align}
 \label{eq:mode_function}
 \psi_n=\frac{1}{\sqrt{2N}}\sum_{q}&e^{\frac{i\pi q n}{N}}\left(\frac{M+(-1)^n(\omega_q-\pi_q)}{\sqrt{2\omega_q(\omega_q-\pi_q)}}a_q\right. \notag \\
               &\left.-\frac{M-(-1)^n(\omega_q+\pi_q)}{\sqrt{2\omega_q(\omega_q+\pi_q)}}\fh{c_{q}^\dagger}\right) \ ,
\end{align}
and the momentum space creation/annihilation operators are defined with respect to the fully filled lower band $\ket{vac}$ according to $a_q\ket{vac}=\fh{c_q}\ket{vac}=0$.
The bosonic samples are prepared in an excited eigenstate of 
\begin{align}
H_{b,0}&=\frac{g^2a_S}{2}\sum_n L^2_{z,n} \,.
\end{align}
determined by the number of bosonic atoms on each site $2\ell=b^\dagger_n b_n +d_n^\dagger d_n$ and the eigenvalue of the operator $L_{z,n}=(b^\dagger_n b-d^\dagger_n d)/2$.
The initial conditions for the bosons need to be chosen such that Gauss's law is fulfilled.
Since the bosonic degrees of freedom are highly occupied, we approximate the initial Wigner function by
\begin{align}\label{Eq:PhiBInit}
\rho_W(\varphi_0) = \prod_n \delta(b_n -\mathcal{B}_{0,n} ) \delta(d_n - \mathcal{D}_{0,n} ) \,.
\end{align}
In fact, the corrections to the exact Wigner function are $\mathcal{O}(1/\mathcal{\ell})$ \cite{2010AnPhy.325.1790P}.
The explicit values of $\mathcal{B}_{0,n}$ and $\mathcal{D}_{0,n}$ are specified in the next section.
To initiate the dynamic evolution, the system is then quenched to an interacting field theory governed by the Hamiltonian \eqref{eq:Hamiltonian}.

\section{Schwinger Pair Production and String Breaking \label{sec:VI}}
In the following we discuss two fundamental phenomena of high-energy physics described by Schwinger model and whose dynamics might be addressed in the  cold-atom framework presented. First we discuss
how the cold atom system approaches the continuum results
of Schwinger pair production and then  we concentrate
on parameter sets that characterize given experimental systems.

\subsection{Theoretical results}
\textit{Schwinger Pair Production.} In quantum electrodynamics the presence of a sufficiently strong electric field results in the spontaneous breakdown of the vacuum by the emission of charged particle-antiparticle pairs (Schwinger effect) \cite{1931ZPhy...69..742S,1936ZPhy...98..714H,1951PhRv...82..664S}.
This fundamental process has not been experimentally observed yet due to the large required field strength of the order of the critical electric field $E_c=M^2/g$.
The observation of this effect in the cold-atom framework, however, seems to be feasible with current technology as discussed in section \ref{sec:IV}.

To study the Schwinger effect in the cold-atom framework, we consider a one-dimensional lattice with $2N$ lattice sites, periodic boundary conditions and finite spin magnitude $\ell=\frac{1}{2}(N_b+N_d)$.
An initially constant electric field $E/E_c = 1$ in QED then corresponds to an initial configuration with a bosonic species imbalance $|I(0)|=|N_b-N_d|=2 M^2/g^2$.
We first solve the equations of motion in the limit $\ell\to\infty$ for $g/M=0.1$, $a_SM=0.005$ and $N=512$.
We checked that our results are insensitive to changes of both the infrared and the ultraviolet cutoff.
 
\begin{figure*}[t!]
\includegraphics{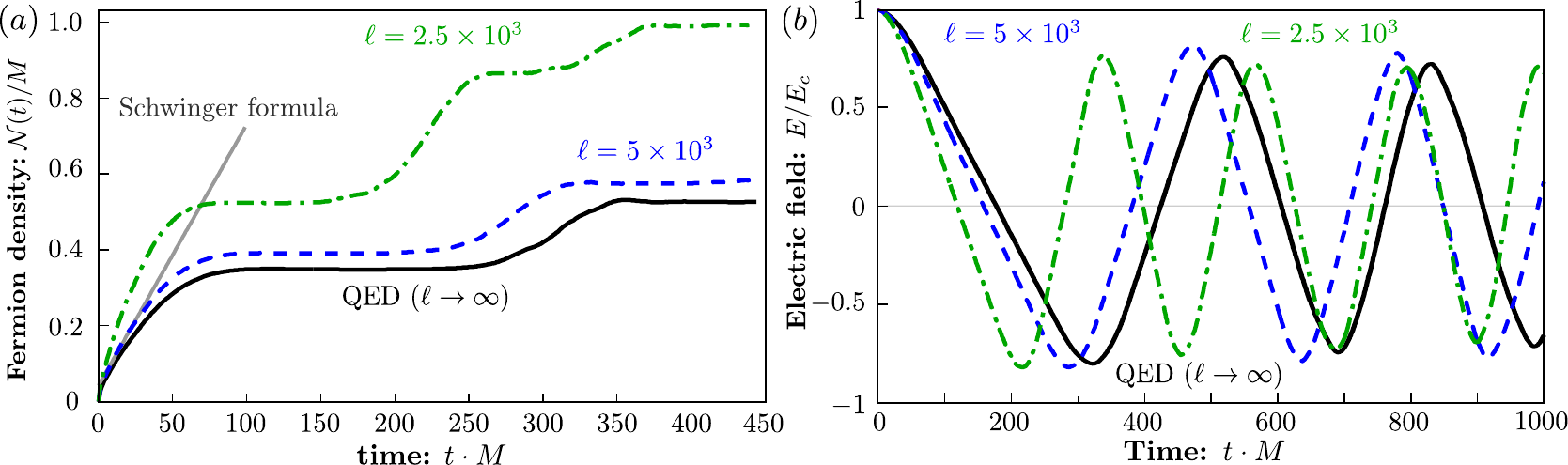}
 \caption{
  Pair production: (a) Time-evolution of the total particle number $\mathcal{N}$ for different values of $\ell$ with fixed $g/M = 0.1$, $a_S M = 0.005$ and $N=512$.
 The dashed line corresponds to the analytic result \eqref{eq:Schwinger_formula}.
 We observe convergence towards the QED result upon increasing the value of $\ell$. (b) Time-evolution of the electric field $E$ for different values of $\ell$ with fixed $g/M = 0.1$, $a_S M = 0.005$ and $N=512$.
 The backreaction of the created particles results in plasma oscillations.
We again observe convergence towards the QED result upon increasing the value of $\ell$.}
 \label{fig:SE_particle_number}
\end{figure*}

In fact, the information of the fermionic sector is encoded in the correlation function $F_{nm}$.
Even though the concept of a particle number is not uniquely defined in an interacting theory \cite{2014PhRvD..90b5021D}, it is useful  to define a quasi-particle distribution $n(k)$ from $F_{nm}$ \cite{2010PhRvD..82j5026H,2013PhRvD..87j5006H,2014PhRvD..90b5016K}, cf.\ also Appendix~\ref{sec:app_particle}.
We display the time-evolution of the total particle number $\mathcal{N}=\sum_k n(k)$ in Fig.~\ref{fig:SE_particle_number}.
The production rate at early times, when the backreaction of produced particles does not yet substantially influence the dynamics, coincides with the analytically known result
\cite{2010PhRvD..82j5026H,2013PhRvD..87j5006H}
\begin{align}
 \label{eq:Schwinger_formula}
 \dot{\mathcal{N}}= \frac{M^2 E}{2\pi E_c} \exp \left(-\pi\frac{E_c}{E}\right) \ . 
\end{align}
At later times, the backreaction of particles becomes important and leads to the expected deviations from the analytic curve.
We find phases in which particle production terminates and plateaus are formed \cite{2013PhRvD..87j5006H, 2014PhRvD..90b5016K}.

%

In the cold-atom setup no fundamental particles are produced since the number of atoms is fixed. 
However, the physics of pair production is still encoded in the correlation function $F_{nm}$ since the staggered structure of the cold-atom setup results in the representation of fermions on even/odd sites as particles/antiparticles. 
Accordingly, the hopping of fermionic atoms between neighboring sites which generates correlations $F_{nm}$ can be interpreted as pair production.
As the cold-atom setup shows a truncation error of $\mathcal{O}(\delta\rho/\ell)$ compared to QED it is interesting to investigate this error as a function of $\ell$.
In Fig.~\ref{fig:SE_particle_number}, we demonstrate the convergence of the cold-atom behavior towards the QED result upon increasing $\ell$. 
For the chosen parameters and $\ell={2500}$, we still observe sizable quantitative deviations from the QED result whereas this discrepancy further decreases for increasing values of $\ell$.

Besides studying fermionic observables based on $F_{nm}$, it is also instructive to evaluate the bosonic species imbalance $I(t)=N_b(t)-N_d(t)$ which is related to the QED electric field according to $E(t)=gI(t)/2$. 
In Fig.~\ref{fig:SE_particle_number}, we display the time-evolution of the electric field.  
Starting from QED ($\ell\to\infty$), we observe the expected behavior according to which the production and subsequent acceleration of particle-antiparticle pairs results in plasma oscillations \cite{2013PhRvD..87j5006H, 2014PhRvD..90b5016K, PhysRevD.45.4659}.
Accordingly, the electric field decreases as the particle number increases and particle creation effectively terminates once the field drops below a certain level, corresponding to the plateaus in the particle number in Fig.~\ref{fig:SE_particle_number}.
 
In the cold-atom setup, the physics of plasma oscillations is observed as well.
The fermionic hopping reduces the initial bosonic species imbalance $I(0)=2 M^2/g^2>0$ until it changes sign and reaches a local minimum $I(t_{\rm{min}})<0$.
Subsequently, the species imbalance increases again, changes sign and reaches a local maximum and so forth.
As for the particle number, we observe that the cold-atom behavior converges towards the QED results upon increasing the value of $\ell$.

\begin{figure}[b!]
 \includegraphics[width=\columnwidth]{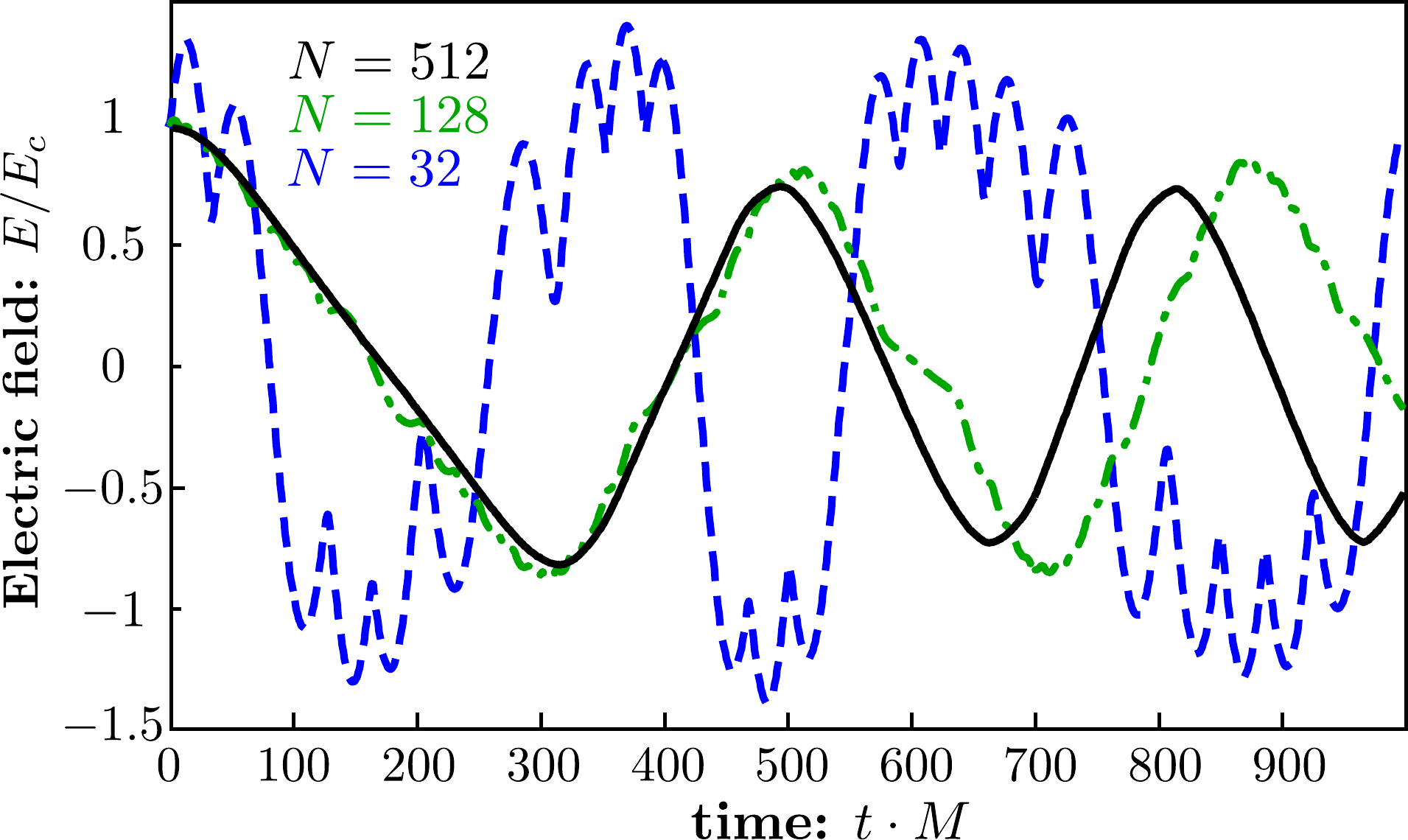}
 \caption{
 Pair production: Time-evolution of the electric field $E$ for different values of $N$ and fixed $g/M = 0.1$, $a_S M = 0.005$ and $\ell=10^5$.}
 \label{fig:SE_finite_size}
\end{figure}

The results in Fig.~\ref{fig:SE_particle_number} are all based on system sizes of $N=512$ for which infrared artifacts are suppressed.
In the following we study the influence of the system size and display the time-evolution of the bosonic species imbalance $I(t)$ for different values of $N$ and fixed $\ell={10^5}$ in Fig.~\ref{fig:SE_finite_size}.
Whereas the behavior remains the same qualitatively, the actual quantitative behavior might substantially change upon decreasing the value of $N$.
For $N$ being too small, we observe oscillations on top of the plasma oscillations which can be attributed to the finite momentum resolution.
Moreover, even though a reasonable agreement between QED and the cold-atom setup is found for the first oscillation period for $N=512$, we still observe sizable deviation at later times.
\newline
\newline
\indent \textit{String Breaking.} The physics of confinement in the theory of quantum chromodynamics (QCD) manifests itself by the formation of a string between two external, static quarks.
This confining string can break in theories with dynamical fermions by the production of charged particle-antiparticle pairs which result in a screening of the static sources \cite{Knechtli1998345, Knechtli2000309, PhysRevLett.81.4056, Gliozzi200591, PhysRevD.71.114513, PhysRevLett.102.191601}.
QED in one spatial dimension shares important
aspect of dynamical string breaking and, therefore, serves as a  toy model for addressing related  questions 
\cite{2013PhRvL.111t1601H, 2014PhRvD..90d5034H,PhysRevD.87.077501,PhysRevD.89.074053,PhysRevD.44.257}.

To study dynamical string breaking in QED in one spatial dimension we prepare two static charges $\pm Q$ located at $\pm d/2$ on the spatial lattice with $2N$ lattice sites.
The corresponding electric field between the charges is given by $E_0=Q$ while it vanishes outside.
In the cold-atom setup, this corresponds to a bosonic species imbalance of $I(0)=2Q/g$ inside the string $|x|<d/2$ whereas it vanishes outside of it.

\begin{figure}[b!]
 \includegraphics[width=0.95\columnwidth]{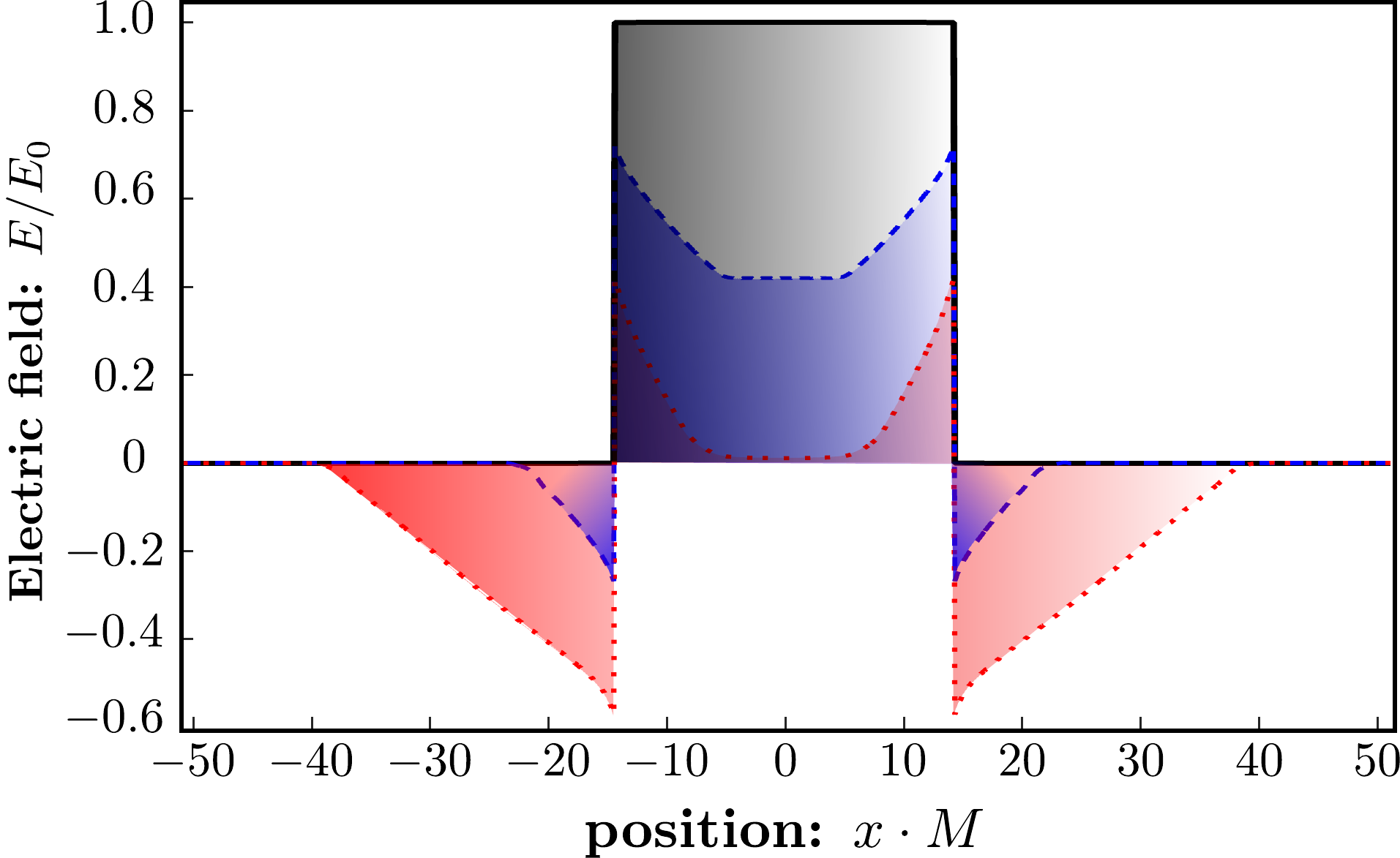}
 \caption{
 String breaking: Electric field $E$ at different times $t_1 \cdot M = 0.0$ (black), $t_2  \cdot M = 8.4$ (blue) and $t_3 \cdot M = 24.9$ (red) for $\ell\to\infty$, $g/M = 1.0$, $a_S M = 0.1$ and $N=1024$.}
 \label{fig:SB_electric_field}
\end{figure}

We first make contact to the corresponding QED literature \cite{2013PhRvL.111t1601H, 2014PhRvD..90d5034H} by considering the limit $\ell\to\infty$ and choosing $g/M=1$, $a_SM=0.1$ and $N=1024$.
To this end, we study the time-evolution of the electric field $E_n$ for $d/a_S=287$ and display different instances of time in Fig.~\ref{fig:SB_electric_field}.
Starting from the initial field configuration, the field energy is transferred to the fermionic sector by particle-antiparticle production such that the amplitude decreases.
The dynamics is such that the opposite charges are produced locally on top of each other and are then accelerated by the electric field.
Depending on the value of $d$, the initial string may or may not contain enough energy to produce the required charges $\pm Q$ to screen the external charges.
In the first scenario the string does not break completely and in the former case
the string starts oscillating.

\begin{figure}[t]
 \includegraphics[width=0.98\columnwidth]{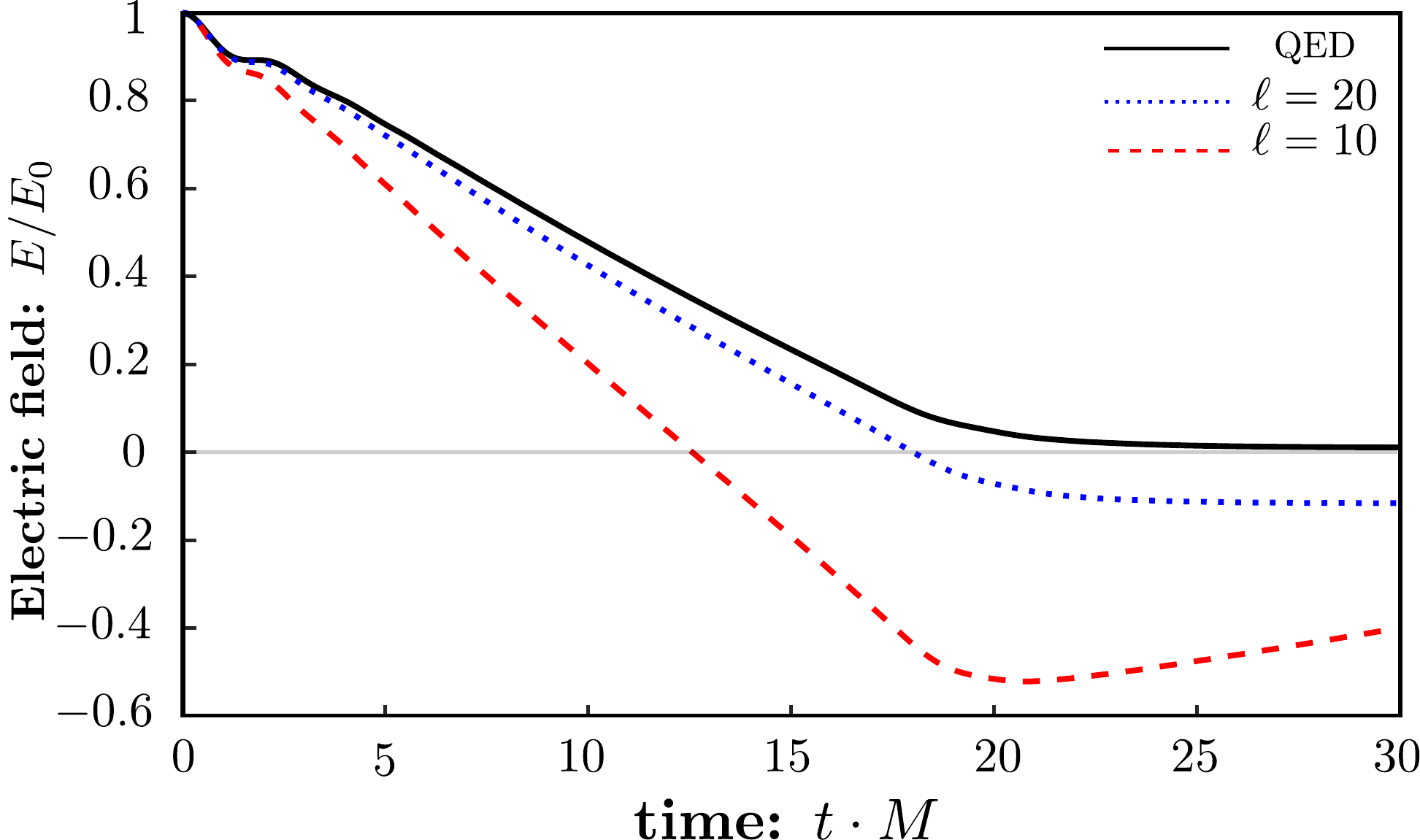}
 \caption{
 String breaking: Time evolution of the electric field $E$ in the center of the string for different values of $\ell$ with fixed $g/M = 1.0$, $a_S M = 0.1$, $N=1024$. The distance between the charges is $d/a_S=287$.
 The zero-crossing of the electric field is attributed to the phenomenon of string breaking.}
 \label{fig:SB_breaking}
\end{figure}

In Fig.~\ref{fig:SB_breaking} we display the electric field in the center of the string, and we choose $d$ such that the produced amount of charge exactly screens the external charges, which is attributed to the phenomenon of string breaking.
Considering the cold-atom setup, the finite value of $\ell$ then again introduces deviations from the QED behavior.
Most notably, we observe that the breaking of the string happens already for smaller distances $d_{CA}<d_{QED}$ for the same parameters $g/M=1$, $a_SM=0.1$ and $N=1024$.
As expected, we observe convergence towards the QED results upon increasing the value of $\ell$.

\begin{figure}[t!]
 \includegraphics[width=\columnwidth]{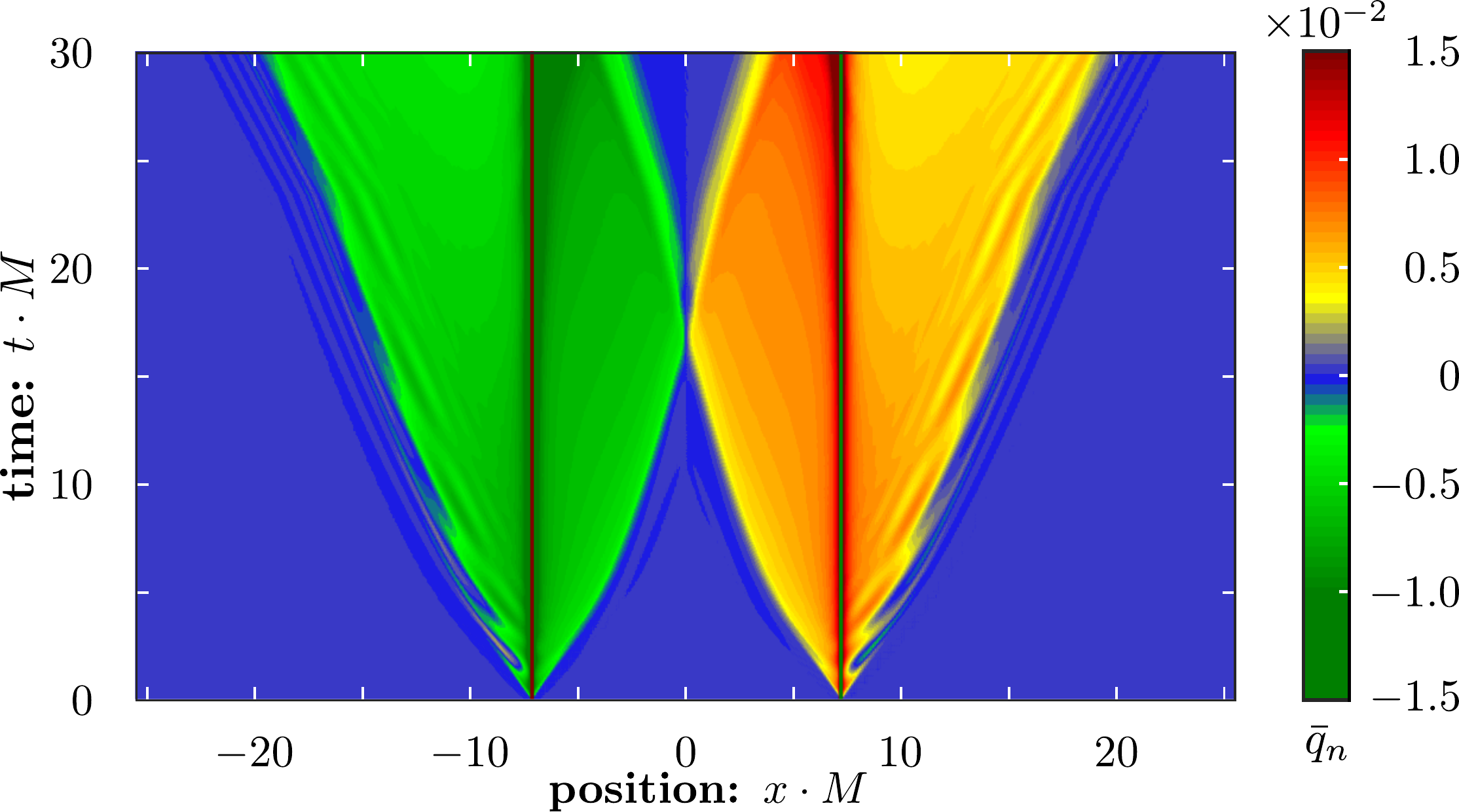}
 \caption{
 String breaking: Time-evolution of the charge density $\bar{q}_n$ for fixed $\ell\rightarrow \infty$, $g/M = 1.0$, $a_S M = 0.1$ and $N=1024$. The distance between the charges is $d/a_S=287$.
 The charges are produced on top of each other and are then separated by the field. 
 Positive charges are accelerated towards $-Q$ whereas the negative charges are accelerated towards $+Q$.}
 \label{fig:SB_charge_density}
\end{figure}

Finally, we consider fermionic observables which are defined in terms of the correlation function $F_{nm}$.
Unlike in the Schwinger mechanism, however, we may observe charge separation directly owing to the spatially inhomogeneous configuration.
Accordingly, we focus on the expectation value of the charge density.
More specifically, we consider the average charge density on two neighboring lattice sites in order to coarse grain the staggered structure, which is an artifact of the chosen fermion discretization
\begin{align}
 \bar{q}_{n}=q_{2n}+q_{2n+1}=-\frac{1}{2}\left(F_{2n,2n}+F_{2n+1,2n+1}\right) \ .
\end{align}
In Fig.~\ref{fig:SB_charge_density} we display the time evolution of the charge density $\bar{q}_n$.
As described previously, the dynamical charges are produced on top of each other such that the total charge density vanishes initially.

The dynamical charges are then separated by the existing field such that positive charges are accelerated towards $-Q$ and negative charge towards $+Q$.
As the dynamical charges cannot be considered as hardcore particles, the charge density spreads beyond the static charges resulting in the outwards directed parts of the charge density.
Accordingly, the external charges are gradually screened and finally result in the breaking of the string.
At asymptotic times, the external charges are then supposed to become screened by an exponential cloud of dynamical fermions \cite{2013PhRvL.111t1601H,2014PhRvD..90d5034H,Pichler-2015, 1990PThPh..84..142I}.

\subsection{Experimental protocol \label{sec:VIB}}

After discussing strong field QED in the continuum limit, we now present an experimental protocol of initialization, evolution and detection that allows us to observe QED with cold atoms.
It is important to note that it does not suffice to only provide the required symmetry of the Hamiltonian in order to quantum simulate a gauge theory but, equally important, also the initial conditions have to fulfill Gauss's law as accurately as possible.
The second requirement can only be guaranteed to a certain degree, however, the presented preparation scheme is consistent with the initial conditions chosen for the theoretical description.

According to section~\ref{Sec:FermInteractions}, the fermions can be prepared in the lowest band via a subsequential adiabatic ramp of $ V^{f}_{2,\alpha}$ and then $ V^{f}_{1,\alpha}$.
If the wavelength of this lattice is well chosen, the bosonic links with $N_B$ atoms are also directly prepared at the intersection between the fermionic sites. 
In particular, the chosen wavelength is supposed to be red detuned for lithium and blue detuned for sodium. 
To apply an initial electric field according to \eqref{Eq:PhiBInit}, the bosons need to be prepared in a staggered structure of alternating imbalance. 
This can be achieved by controlling the imbalance with a linear coupling between the two bosonic states, e.g., via radio-frequency coupling or two photon microwave coupling. 
The detuning of the linear coupling for every second site is controlled by utilizing a species selective standing light wave with twice the lattice period for the bosons.
Properly chosen, one can implement a $\pi$ pulse for every second site that leads to the required 'staggered' imbalance of the bosons. 
\fh{The subsequent dynamics of the system is initiated with a quench of the mass term from being far off-resonant.}

\begin{figure*}[t!]
 \includegraphics[width=2\columnwidth]{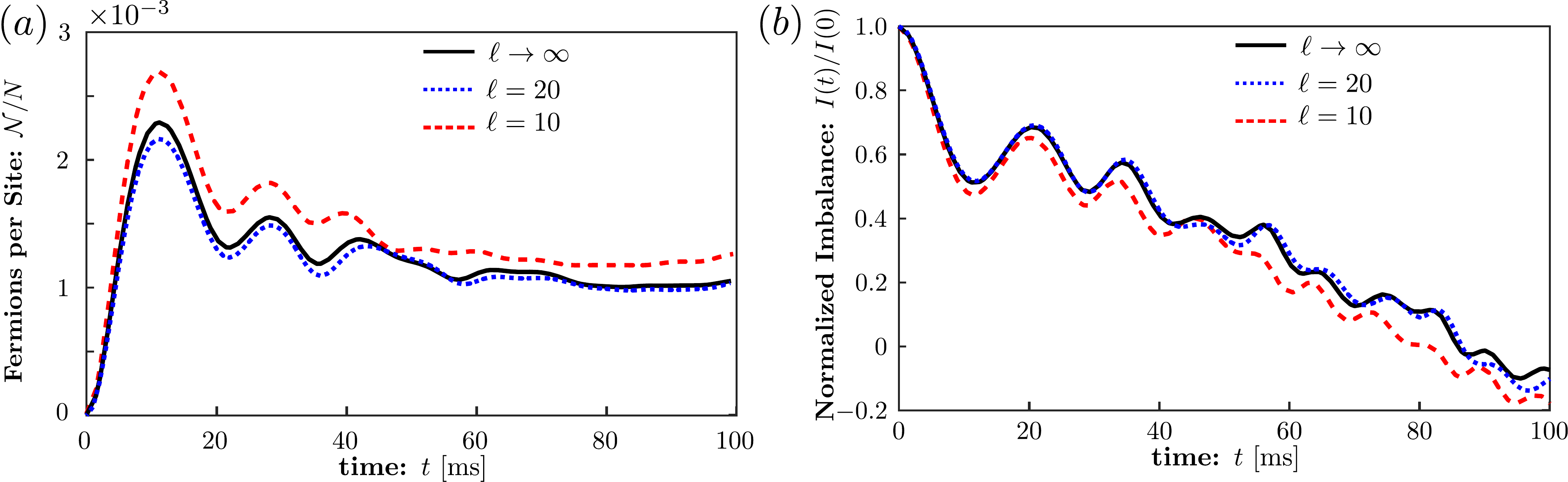}
 \caption{ 
 Particle production in $^{23}$Na-$^6$Li setup: The initial production of particles is driven by $\chi_{BF}N_B$ with an initial imbalance of $I(0) = 0.3$, which subsequently levels off so that the particle number reaches a plateau. 
 We observe quick convergence towards QED as function of $\ell$, where the parameters for this simulation are $g/M = 2.6$, $a_S M = 0.05$ and $N=100$.}
 \label{fig:NaLiParticleProduction}
\end{figure*}

We start our benchmarking with the parameters presented in section~\ref{sec:IV}, which correspond to $g/M=2.6$,  $a_S M=0.05$ and $N=100$.
For these initial conditions we can benchmark a possible experimental realization via the functional integral approach.
This allows us to investigate the role of the experimentally relevant parameters on the physics of the Schwinger effect, especially the spin magnitude $\ell=N_B/2$ and the coupling strength $g$.
To check the convergence towards the lattice QED result for given parameters we also vary the spin magnitude, $\ell=10,20,\infty$.
Since the experimental parameter allow for the exploration of the strong coupling regime $g/M>1$, the range of validity of the theoretical treatment has to be further investigated \cite{2016arXiv161200739B}.
We expect, however, that the experiment shows the same qualitative behavior as shown in Fig.~\ref{fig:NaLiParticleProduction}.

In Fig.~\ref{fig:NaLiParticleProduction}a we display the time evolution of particle number per lattice site, where we use again the adiabatic definition from appendix~\ref{sec:app_particle}.
First, we observe that particle production happens initially on time scales of the order of $\chi_{BF}N_B$, which is short compared to the limiting particles losses, and thus accessible in the experiment. 
As to be expected from $\chi_{BF}N_B>\Delta$ we also observe initial oscillations, which are smeared on times $t>h/\Delta$. 
The particle number per lattice site then reaches a plateau with $\mathcal{N}(t)/N\simeq10^{-3}$.
For the given parameters, we find convergence towards the QED results already for $\ell=20$ which, as compared to the ideal-typical parameters from the previous section, can be traced back to the larger value of the coupling $g/M=2.6$.
In general it turns out that the required value of $\ell$ is inversely proportional to $g/M$ to obtain convergence towards the QED result, cf. also the simulations of dynamical string breaking in Fig.~\ref{fig:SB_breaking}.
Owing to the fact that realistic values of the spin magnitude are of the order of $\ell=\mathcal{O}(100)$, we can expect genuine QED behavior for the proposed experimental setup.
\fh{The fluctuations in the number of bosons, which is expected to be of the order of a few percent, are not expected to substantially alter the reported behavior on the time scales considered.}
We also display the electric field $E(t)=g I(t)/2$ as determined by the species imbalance $I(t)=N_b(t)-N_d(t)$ in Fig.~\ref{fig:NaLiParticleProduction}b.
The production of particles results in a decrease of the species imbalance, which quickly drops below the critical value so that particle production terminates.

Conceptually the momentum distribution of the produced particles can be read out precisely via band mapping \cite{Scelle2013a}, however, this is very challenging for the given parameters as one can deduce from Fig.~\ref{fig:NaLiParticleProduction}. Nevertheless, the underlying physics of particle production can be already accessed from the integrated number of produced particles.
Nevertheless, the underlying physics of particle production can be already accessed from the integrated number of produced particles.
According to Fig.~\ref{fig:NaLiParticleProduction}, around $0.2$ particles have to be detected on average for a lattice with $N=100$ sites.
Current experimental setups allow realizing $\approx 30$ copies of the cold atom QED system by employing an array of one dimensional traps.
Thus one expects integrated $\mathcal{O}(10)$ particles which can be detected with fluorescence in a subsequent magneto-optical trap for which single particle resolution is well established \cite{Hume-2013}.
While the detection of the produced particles is very challenging, the bosonic species imbalance, i.e. the electric field, changes significantly.
Thus, the Schwinger effect in a cold atomic setup can also be observed by measuring the integrated boson imbalance via standard absorption imaging techniques.

\begin{figure}[b!]
 \includegraphics[width=0.96\columnwidth]{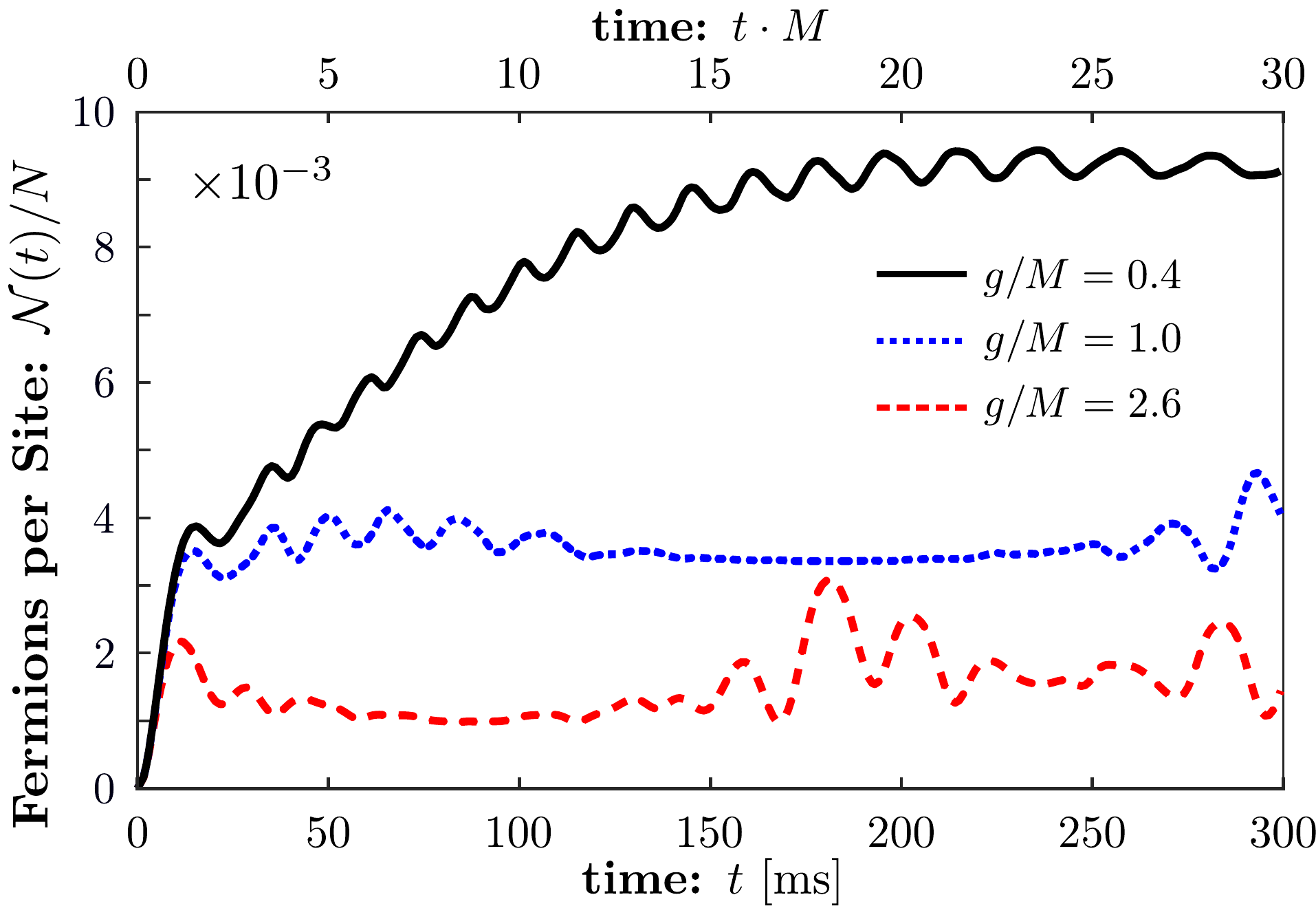}
 \caption{
 Coupling dependence in $^{23}$Na-$^6$Li setup: Time-evolution of the particle number for different $g/M=0.4,1.0,2.6$ and fixed $a_S M =0.05$, $\ell=100$, $N=100$. 
 We adapt the initial species imbalance to provide for a critical initial field $E_{\text{cr}}$ in each case.}
 \label{fig:Coupling}
\end{figure}

The coupling strength $g/M$ is another experimentally relevant parameter of importance.
Accordingly, we study the dependence of the particle number on choosing slightly different couplings $g/M=0.4,1.0,2.6$ for fixed parameters $a_S M=0.05$, $N=100$ and $\ell\to\infty$ in Fig.~\ref{fig:Coupling}.
We note that we have to adapt the initial species imbalance $I(0)=2 M^2/g^2$ in order to provide an initial critical field $E_{\text{cr}}$ in each case.  
Most notably, the simulations indicate that the first plateau in the particle number is a stable signature for the non-trivial interplay of the electric field and the produced fermions.
We find, however, that the number of produced particles at the first plateau is inversely proportional to the ratio $g/M$.
As expected, smaller couplings $g/M$ result in a slowdown of dynamics such that the first plateau is reached at later times.

\section{Conclusion \label{sec:VII}}

This work is meant as a guide towards a first large-scale quantum simulation of a lattice gauge theory with dynamical gauge fields. 
For that purpose, we concentrate on one of the simplest, yet highly nontrivial, example of QED in 1+1 space-time dimensions and provide strong evidence that present-day experimental resources and protocols are sufficient to observe the dynamical phenomena of Schwinger pair production and string breaking in the laboratory using ultracold atoms. 

Our results point out that experimental realizations using coherent many-body states residing on the links of an optical lattice can be highly efficient for quantum simulations of such high-energy particle physics phenomena. 
This represents a paradigmatic change in view of the large number of studies in the literature focusing on a small number of atoms per link. 
To substantiate our findings, we exploit the long-term experience that has been gained with the engineering and manipulation of related setups of fermions interacting with coherent samples of bosonic atoms. 

For the example of a Bose-Fermi mixture of $^{23}$Na and $^6$Li atoms, characterized by a plethora of potentially gauge-symmetry breaking interactions, we apply external potentials and fields such that one ends up with a lattice Hamiltonian with local gauge invariance. 
In this way, the microscopic parameters describing the bosonic and fermionic atom degrees of freedom are connected with the parameters describing the gauge field theory.

We use the experimentally available parameter range of the cold-atom system in benchmark calculations and convergence against the full QED results is observed. 
The very detailed comparisons of the real-time dynamics of the atomic system in the large boson number regime are possible using powerful functional integral techniques, which complement exact diagonalization or tensor network methods that are applicable in the small boson number regime. 

A future experimental implementation of gauge symmetries in a flexible cold-atom setup can explore new parameter ranges and phenomena, even beyond what is realized by nature so far. 
While the gauge coupling of QED is weak, with $\alpha_{\rm em} \simeq 1/137$ in nature's three spatial dimensions, studies at stronger couplings in various dimensions would be extremely interesting.
No conventional computational technique has so far been able to predict the real-time dynamics of QED or QCD for couplings of order one -- despite the fact that this is a crucial missing link in our understanding of the thermalization process of QCD as explored in relativistic heavy-ion collision experiments. 
The experimental setup discussed in this work may provide already access to answering important aspects of longstanding open questions.    

\begin{acknowledgments}
We thank M.~Dalmonte, E.~Demler, A.~Frishman, T.~Gasenzer, J.~G\"oltz, M.~Karl, N.~M\"uller, J.~Pawlowski, A.~Polkovnikov, I.~Rocca and U.-J.~Wiese for helpful discussions and collaborations on related work.
V.~Kasper was supported by the Max Planck society.
F.~Hebenstreit acknowledges support from the European Research Council under the European Union's Seventh Framework Programme (FP7/2007-2013)/ ERC grant agreement 339220. 
This work is part of and supported by the DFG Collaborative Research Centre "SFB \fh{1225} (ISOQUANT)".
\end{acknowledgments}

\appendix

\section{Overlap integrals \label{sec:app_overlap}}

In this appendix, we determine the overlap integrals which appear in the main text.
To this end, we consider an optical lattice with lattice constant $a$ and tight radial confinement so that we assume that the radial and longitudinal directions decouple. 
Further, the bosonic and fermionic atoms are supposed to be in the ground state with respect to the radial direction. 
Assuming a harmonic potential, the ground state wave functions introduced in \eqref{eq:Reduction} are given by
\begin{align}
 \label{eq:app_radial}
 \varphi_s(y)=\left(\pi a^2_{s,\perp}\right)^{-1/4}e^{-\frac{1}{2}\left(\frac{y}{a_{s,\perp}}\right)^2} \ ,
\end{align} 
with the harmonic oscillator length scale
\begin{align}
 a_{s,\perp} = \sqrt{\frac{\hbar}{M_s\omega_{s,\perp}}} \, .
\end{align}
The expressions for $\varphi_{s}(z)$ follow from replacing $y$ by $z$.
Here, we assumed that the ground state wave functions are independent of the magnetic quantum number $\alpha$.

In the longitudinal direction, the optical lattice potential is determined by 
\begin{subequations}
\begin{align}
V^{b}_{\parallel}(x) &=  V^{b}_{1} \cos^2 (2k x)  \ , \\
V^{f}_{\parallel}(x) &=  V^{f}_{1} \sin^2 (2k x) + V^{f}_{2} \cos^2 (k x) \ ,
\end{align}
\end{subequations}
with $k=\pi/a$ and $V^{b}_1>0$.
The locations of the minima of the bosonic potential are at
\begin{subequations}
\begin{align}
 x_{b,n} &= \frac{2n+1}{4}a  \, .
\end{align}
\end{subequations}
In the vicinity of the potential minimum $x_{b,n}$, we may expand the potential in a Taylor series to quadratic order
\begin{align}
V^b_\parallel(x) \approx  4 k^2 V_1^b (x-x_{b,n})^2 \ .
\end{align}
The minima in the fermionic potential are determined by
\begin{align}
 x_{f,n} &= \frac{n}{2}a  \, .
\end{align}
For the fermionic atoms, there are two different types of local minima and we need to expand the optical potential in a Taylor series to quadratic order around both of them.
For even sites, such as $x_{f,0}=0$, and odd sites sites, such as $x_{f,1}= \frac{1}{2}a$, we obtain
\begin{subequations}
 \begin{align}
  V^f_\parallel(x)|_{x_{f,0}} &\approx V^f_2 + (4V^f_1 -V_1^f )k^2 (x-x_{f,0})^2 \, , \\
  V^f_\parallel(x)|_{x_{f,1}} &\approx (4V^f_1 +V^f_2) k^2(x-x_{f,1})^2\, .
 \end{align}
\end{subequations}
The quadratic terms in the potential expansions determine the harmonic oscillator frequencies
\begin{subequations}
\begin{align}
\frac{1}{2}M_b\omega_{b,\parallel}^2 &= 4 k^2 V_1^b \, ,\\
\frac{1}{2}M_f\omega_{f,\parallel, L}^2 &= (4V^f_1 -V_1^f )k^2 \, ,\\
\frac{1}{2}M_f\omega_{f,\parallel, R}^2 &= (4V^f_1 +V^f_2) k^2\, .
\end{align}
\end{subequations}
We require $4V^f_1-V^f_2>0$ such that the oscillator frequencies are real.
Note again that we have two different frequencies for the fermions corresponding to even/odd sites whereas there is only a single bosonic frequency.
The corresponding oscillator length scales are then given by
\begin{subequations}
\begin{align}
a_{b,\parallel} &=  \sqrt{\frac{\hbar}{M_b\omega_{b,\parallel}}}\, , \\
a_{f,\parallel, p} &= \sqrt{\frac{\hbar}{M_f\omega_{f,\parallel, p}}} \ ,
\end{align}
\end{subequations}
with $p=L,R$. 
The corresponding Wannier functions are 
\begin{subequations}
\label{eq:app_perp}
\begin{align}
 w^b_{n}(x) &=  (\pi a^2_{b,\parallel})^{-1/4} e^{-\tfrac{1}{2} \left( \frac{x-x_{b,n}}{a_{b,\parallel}} \right)^2 } \,, \\
 w^f_{n}(x) &=  (\pi a^2_{f,\parallel})^{-1/4} e^{-\tfrac{1}{2}\left( \frac{x-x_{f,n}}{a_{f,\parallel, L}} \right)^2} \,, 
\end{align}
\end{subequations}
where we assumed again that the wave functions are independent of the magnetic quantum number, i.e.\  $w^b_{n}(x) =  w^b_{\alpha,n}(x)$ and $w^f_{n}(x) = w^f_{\alpha,n}(x)$.

Upon performing the dimensional reduction and change of basis to Wannier function, the following overlap integrals appear in the interaction terms:
\begin{subequations}
\label{eq:Overlaps}
\begin{align}
U^b_{\mathbf{n}}  &= \int dy  \, |\varphi_{b}(y)|^4 
 \int dz  |\varphi_{b}(z)|^4 \notag \\
&\times \int dx \, [ w^{b}_{n_1}(x)w^{b}_{n_2}(x)]^{\ast}w^b_{n_3}(x)w^b_{n_4}(x) \\
U^f_{\mathbf{n}}  &= \int dy  \, |\varphi_{f}(y)|^4 
 \int dz  |\varphi_{f}(z)|^4 \notag \\
&\times \int dx \, [ w^{f}_{n_1}(x)w^{f}_{n_2}(x)]^{\ast}w^f_{n_3}(x)w^f_{n_4}(x) \\
U^{bf}_{\mathbf{n}}  &=  \int dy  \, |\varphi_{b}(y)\varphi_{f}(y)|^2
 \int dz  |\varphi_{b}(z)\varphi_{f}(y)|^2 \notag \\
& \times\int dx \,  [ w^{f}_{n_1}(x)w^{b}_{n_2}(x)]^{\ast}w^b_{n_3}(x)w^f_{n_4}(x)
\, .
\end{align}
\end{subequations}
These are Gaussian integrals, which can be determined analytically, cf. \eqref{eq:app_radial} and \eqref{eq:app_perp}.
Further we the bosonic and fermionic tunnel elements
\begin{subequations}
\label{eq:OverlapIntegral}
\begin{align}
\! \!\!\!\!J_B &= \int dx \, w^b_n(x)\left(-\frac{\hbar^2}{2M_b} \frac{\partial^2}{\partial x^2 } + V^b(x)   \right) w^b_{n+1}(x) \, ,\\
\! \!\!\!\!J_F &= \int dx\,  w^f_n(x)\left(-\frac{\hbar^2}{2M_f}\frac{\partial^2}{\partial x^2 } 
 + V^f(x)  \right)w^b_{n+1}(x) \,,
\end{align} 
\end{subequations}
are used in section \ref{sec:IV}. 

\section{Coherent states and Wigner transform \label{sec:app_wigner}}
In this appendix we summarize the main definitions that are used in section~\ref{sec:V} (see also Ref.~$\!$\cite{1969PhRv..177.1882C}).
The coherent state of a single bosonic mode 
\begin{align}
 \ket{\varphi}=e^{-\frac{1}{2}|\varphi|^2}e^{\varphi \phi^\dagger}\ket{0}=D(\varphi)\ket{0} \ , 
\end{align}
is defined as the right eigenstate of the bosonic annihilation operator, $\phi\ket{\varphi}=\varphi\ket{\varphi}$, where we introduced the displacement operator
\begin{align}
 D(\varphi) = \exp(\varphi \phi^{\dagger} - \varphi^* \phi) \ .
\end{align}
The identity operator reads
\begin{align}
 \mathbbm{1}=\iint{\frac{d\Re{\varphi}\,d\Im{\varphi}}{\pi}\ket{\varphi}\!\bra{\varphi}}\equiv \int{\frac{d^2\varphi}{\pi}\ket{\varphi}\!\bra{\varphi}} \ .
\end{align}
The complex Dirac-delta function
\begin{align}
 \label{eq:app_delta}
 \delta(\varphi_1-\varphi_2)\equiv\delta(\Re{\varphi_1}-\Re{\varphi_2})\delta(\Im{\varphi_1}-\Im{\varphi_2}) 
\end{align}
has the integral representation
\begin{align}
 \int{\frac{d^2\varphi}{\pi}e^{\varphi^*\lambda-\varphi\lambda^*}}=\pi\delta(\lambda) \ .
\end{align}
The Wigner transform of an operator $O(\phi,\phi^\dagger)$ is given by
\begin{align}
\label{eq:DefWignerTrafo}
 O_W(\varphi) = \int \frac{d^2\lambda}{\pi} \Tr \{ O  D^{\dagger}(\lambda)\} e^{\lambda \varphi^{\ast} - \varphi \lambda^{\ast}} \ .
\end{align}
For $O = A B$  with two observables $A(\phi,\phi^\dagger)$ and $B(\phi,\phi^\dagger)$ the Wigner transforms   can be expressed as
\begin{align}
(&AB)_W(\varphi) = \notag \\
\!&\int \frac{d^2 \lambda}{\pi} \int \frac{d^2\eta}{\pi}  e^{\eta(\varphi  -\lambda)^{\ast} - \eta^{\ast}(\varphi  -\lambda)}A_W(\lambda) B_W(\varphi - \tfrac{1}{2}\eta) = \notag \\
\!&\int \frac{d^2 \lambda}{\pi} \int \frac{d^2\eta}{\pi}  e^{\eta(\varphi  -\lambda)^{\ast} - \eta^{\ast}(\varphi  -\lambda)}A_W(\varphi+\tfrac{1}{2}\eta) B_W(\lambda) \ . \label{eq:app_wignerprod}
\end{align}

\section{Particle number distribution  \label{sec:app_particle}}

In this appendix, we define and determine the particle number distribution $n(q)$ in terms of the correlation function $F_{nm}=\bra{vac}[\psi_n,\psi_m^\dagger]\ket{vac}$.
To this end, we consider the fermionic part of the Kogut-Susskind Hamiltonian \eqref{eq:KS_Hamiltonian}
\begin{align}
 \label{eq:app_KS_Hamiltonian}
 H&=\sum_{n=0}^{2N-1} \frac{i(\psi_{n+1}^{\dagger} U^\dagger_n \psi_{n} - \psi_{n}^\dagger U_{n}\psi_{n+1})}{2a_S} + M(-1)^n\psi_n^\dagger \psi_n\ .
\end{align}
We may diagonalize the Hamiltonian by treating the link variables $U_n$ as c-number background for the fermions, as it is also done in the functional integral approach.
To this end, we define the Fourier transformation according to
\begin{subequations}
\label{eq:app_FT_fermions}
\begin{align}
 \psi_n&=\frac{1}{\sqrt{2N}}\sum_{q=0}^{2N-1}e^{\frac{i \pi q n}{N}}\tilde{\psi}_q \ , \\
 \tilde{\psi}_q&=\frac{1}{\sqrt{2N}}\sum_{n=0}^{2N-1}e^{-\frac{i \pi q n}{N}}\psi_n \ .
\end{align}
\end{subequations}
We note that the system is still translation invariant over two lattice sites if we study Schwinger pair production, i.e.\ $U_n=U_{n+2l}$ with $l\in\{0,\ldots,N-1\}$.
Denoting the even links by $U_{\text{even}}=U_{2n}$ and the odd links by $U_{\text{odd}}=U_{2n+1}$, we have
\begin{align}
 U_n={U}_A+(-1)^n{U}_B 
\end{align}
with ${U}_{A}=(U_{\text{even}}+U_{\text{odd}})/2$ and ${U}_{B}=(U_{\text{even}}-U_{\text{odd}})/2$.
The Hamiltonian can then be written in matrix notation according to
\begin{align}
  H=\sum_{q=0}^{N-1}\begin{pmatrix}\tilde{\psi}^\dagger_q&\tilde{\psi}^\dagger_{q+N}\end{pmatrix}\begin{pmatrix} \pi_q&M_q\\M^*_q&-\pi_q\end{pmatrix}\begin{pmatrix}\tilde{\psi}_q\\\tilde{\psi}_{q+N}\end{pmatrix}
\end{align}
with
\begin{subequations}
\begin{align}
 \pi_q&=\frac{i}{2a_S}\left(e^{-\frac{i \pi q}{N}}{U}^*_A-e^{\frac{i \pi q}{N}}{U}_A\right)=\pi_q^* \ , \\
 M_q&=M+\frac{i}{2a_S}\left(e^{-\frac{i \pi q}{N}}{U}^*_B+e^{\frac{i \pi q}{N}}{U}_B\right) \ .
\end{align}
\end{subequations}
Fermions without a gauge field correspond to $U_n=1$ leading to ${U}_A=1$ and ${U}_B=0$.
The eigenvalues of this Hamiltonian are given by $\omega_{\pm,q}=\pm\omega_q$ with
\begin{align}
 \omega_q=\sqrt{M^2+\frac{1}{4a_S^2}\left|e^{\frac{i \pi q}{N}}U_{\text{even}}-e^{-\frac{i \pi q}{N}}U^*_{\text{odd}}\right|^2} \ .
\end{align}
The corresponding normalized eigenvectors are
\begin{subequations}
\label{eq:app_eigenvectors}
\begin{align}
 u_{q}^{+}&=\frac{1}{\sqrt{2\omega_q(\omega_q-\pi_q)}}\begin{pmatrix}M_q\\\omega_q-\pi_q\end{pmatrix} \ , \\
 u_{q}^{-}&=\frac{1}{\sqrt{2\omega_q(\omega_q+\pi_q)}}\begin{pmatrix}-M_q\\\omega_q+\pi_q\end{pmatrix} \ ,
 \end{align}
\end{subequations}
In fact, every $q$-mode is diagonalized by a unitary transformation matrix $U_q=(u^+_{q},u^{-}_{q})$.
This defines quasi-particle creation/annihilation operators
\begin{align}
 \begin{pmatrix} a_q \\ \fh{c^\dagger_{q}}\end{pmatrix}= U^\dagger_q\begin{pmatrix}\tilde{\psi}_q \\ \tilde{\psi}_{q+N} \end{pmatrix} 
 \label{eq:BogTrafo}
\end{align}
with respect to the instantaneous vacuum state $\ket{\Omega}$, fulfilling $a_q\ket{\Omega}=\fh{c_q}\ket{\Omega}=0$.
The Hamiltonian is then given by
\begin{align}
 H=\sum_{q=0}^{N-1}\omega_q(a^\dagger_qa_q+\fh{c^\dagger_qc_q}-1) \ .
\end{align}
We define the quasi-particle distribution function $n(q)$ as the expectation value of the instantaneous number operator 
\begin{align}
 n(q)\equiv \bra{vac}a^\dagger_q a_q+ \fh{c^\dagger_q c_q}\ket{vac} \, ,
\end{align}
where the asymptotic ground state $\ket{vac}$ is given by
the Hamiltonian with $U_n = 1$.
The expectation value of the Hamiltonian is 
\begin{align}
 \mathcal{E}=\bra{vac}H\ket{vac}=\sum_{q=0}^{N-1}\omega_q[n(q)-1] \ .
\end{align}
The two contributions of $n(q)$ are found by employing the Bogoliubov transformation \eqref{eq:BogTrafo} such that
\begin{align}
\!\bra{vac}a^\dagger_qa_q\ket{vac} &= \frac{|M_q|^2}{2\omega_q(\omega_q - \pi_q)} 
\braket{vac|\psi^{\dagger}_q \psi_q|vac} \notag \\
&+\frac{M_q}{2\omega_q} 
\braket{vac|\psi^{\dagger}_q \psi_{q+N}|vac} \notag \\
&+\frac{M^{\ast}_q}{2\omega_q} 
\braket{vac|\psi^{\dagger}_{q+N} \psi_{q}|vac} \notag \\
&+\frac{\omega_q - \pi_q}{2\omega_q} 
\braket{vac|\psi^{\dagger}_{q+N} \psi_{q+N}|vac} ,
\end{align}
and 
\begin{align}
\!\bra{vac} \fh{c^\dagger_q c_q}\ket{vac} &= \frac{|M_q|^2}{2\omega_q(\omega_q + \pi_q)} 
\braket{vac| \psi_q \psi^{\dagger}_q|vac} \notag \\
&-\frac{M^{\ast}_q}{2\omega_q} 
\braket{vac| \psi_q \psi^{\dagger}_{q+N}|vac} \notag \\
&-\frac{M_q}{2\omega_q} 
\braket{vac|\psi_{q+N} \psi^{\dagger}_{q}|vac} \notag \\
&+\frac{\omega_q + \pi_q}{2\omega_q} 
\braket{vac|\psi_{q+N} \psi^{\dagger}_{q+N}|vac} \ .
\end{align}
The Fourier transformation of the correlation function $F_{mn}$ is determined by
\begin{align}
 \label{eq:app_FT_inverse_correlator}
 \tilde{F}_{q,q'}=\bra{vac}[\tilde{\psi}_{q},\tilde{\psi}^\dagger_{q'}]\ket{vac}=\frac{1}{2N}\sum_{n,m=0}^{2N-1}e^{-\frac{i\pi (qn-q'm)}{N}}F_{nm}
\end{align}
with $q,q'\in\{0,\ldots,2N-1\}$.
Accordingly, $n(q)$ can be written as 
\begin{align}
 n(q)=\frac{\varepsilon_q}{\omega_q}+1 \ ,
\end{align}
where the energy density in Fourier space is given by 
\begin{align}
 \varepsilon_q = \frac{1}{2}\left[\pi_q(\tilde{F}_{q+N,q+N}-\tilde{F}_{q,q})-M_q \tilde{F}_{q+N,q}-M^{\ast}_q \tilde{F}_{q,q+N}\right] \, ,
\end{align}
The total particle number is then found by summing over all Fourier modes
\begin{align}
 \mathcal{N}=\sum_{q}n(q) \ .
\end{align}
In the cold atom system, we proceed analogously but replace the link operators by the Schwinger bosons $U_n\to[\ell(\ell+1)]^{-1/2}b^*_nd_n$.
As the bosonic degrees of freedom are again considered as c-numbers, we obtain very similar expressions for the particle number.

\newpage

\bibliographystyle{apsrev4-1}
\bibliography{publications}

\end{document}